%% file: manuscript_elsevier.tex
\documentclass[a4paper,fleqn]{cas-sc}

\usepackage[numbers, square, sort&compress]{natbib}
\usepackage{graphicx}%
\usepackage{multirow}%
\usepackage{amsmath,amssymb,amsfonts}%
\usepackage{amsthm}%
\usepackage{mathrsfs}%
\usepackage[title]{appendix}%
\usepackage{xcolor}%
\usepackage{textcomp}%
\usepackage{manyfoot}%
\usepackage{booktabs}%
\usepackage{algorithm}%
\usepackage{algorithmicx}%
\usepackage{algpseudocode}%
\usepackage{listings}%
\usepackage{tabularx}%
\usepackage{lscape}%
\usepackage{float}
\usepackage{pdflscape}
\usepackage{longtable}
\usepackage{multirow}
\usepackage[utf8]{inputenc}
\usepackage{url}
\usepackage{subcaption}

\def\tsc#1{\csdef{#1}{\textsc{\lowercase{#1}}\xspace}}
\tsc{WGM}
\tsc{QE}
\tsc{EP}
\tsc{PMS}
\tsc{BEC}
\tsc{DE}

\begin{document}
\let\WriteBookmarks\relax
\def\floatpagepagefraction{1}
\def\textpagefraction{.001}
\shorttitle{Non-light water reactors}

\shortauthors{Wimmers et al.}

\title [mode = title]{Can They Compete? Cost Competitiveness of Non-Light-Water Reactors for Heat and Power Supply in a Decarbonized European Energy System}

\author[1,2]{Alexander Wimmers}[orcid=0000-0003-3686-0793]
\cormark[1]
\ead{awi@wip.tu-berlin.de}
\author[1]{Fanny Böse}[orcid=0009-0004-8814-8946]
\author[3]{Leonard Göke}[orcid=0000-0002-3219-7587]

\address[1]{Workgroup for Infrastructure Policy (WIP), Technische Universität Berlin, Straße des 17. Juni 135, 10623 Berlin, Germany}
\address[2]{Energy, Transportation, Environment Department, German Institute for Economic Research (DIW Berlin), Anton-Wilhelm-Arno-Straße 58, 10117 Berlin, Germany}
\address[3]{Reliability and Risk Engineering, ETH Zürich, 8092 Zürich, Switzerland}

\begin{abstract}
Recent pledges to triple global nuclear capacity by 2050 suggest a "nuclear renaissance," bolstered by unconventional reactor concepts such as sodium-cooled fast reactors, high-temperature reactors, and molten salt reactors. These technologies claim to address the challenges of today’s high-capacity light-water reactors, i.e., cost overruns, delays, and social acceptance, while also offering additional non-electrical applications. However, this analysis reveals that none of these concepts currently meet the prerequisites of affordability, competitiveness, or commercial availability. Our cost analysis reveals optimistic FOAK cost assumptions of 5,623 to 9,511 USD per kW, and NOAK cost projections as low as 1,476 USD per kW. At FOAK cost, the applied energy system model for Europe in 2040 includes no nuclear power capacity, and thus indicates that significant cost reductions would be required for these technologies to contribute to energy system decarbonization. In lower-cost scenarios, reactors capable of producing high and medium temperature heat compete with other technologies and dominate the system once costs fall below 5,000 USD per kW. Electricity shares reach current levels of $\approx 20\%$ once costs are reduced to 3,000 USD per kW or less We conclude that, for reactor capacities to increase significantly, a focus on certain technology lines and streamlined regulation in necessary. Further remaining technological challenges, e.g., new waste streams, must be resolved.
\end{abstract}


\begin{highlights}
\item Analysis of historical and future non-electrical applications of nuclear power
\item Detailed cost analysis for the proposed non-light water reactor concepts
\item Energy system modeling of nuclear cogeneration competing with renewables
\end{highlights}



\maketitle 
    \include{contents}
\end{document}

%% file: contents.tex
\section{Introduction}

There is a broad consensus that decarbonizing energy systems is necessary to limit global warming and reduce the potentially hazardous effects of climate change \cite{byers_ar6_2022}. However, extensive debates exist on the means to this end, especially regarding the technology mix for a cost-efficient and executable solution. In these debates, nuclear power is often considered as one of the low-carbon technologies that will play a significant part in global energy system decarbonization \cite{duan_stylized_2022,byers_ar6_2022, ec_commission_2025}, albeit existing uncertainties regarding the cost-efficiency and availability of this technology \cite{rothwell_projected_2022, lovins_us_2022, steigerwald_uncertainties_2023,wealer_investing_2021,bose_questioning_2024, goke_flexible_2025}. In the last decade, these prospects have launched a narrative of a "nuclear renaissance" coinciding with the promotion of so-called novel reactor concepts, or "advanced" reactor technologies, mostly based on non-light water designs \cite{thomas_competitive_2010, nuttall_nuclear_2022}. This is supported by recent governmental proclamations to triple the globally installed nuclear power plant capacity by 2050 \cite{wnn_ministerial_2023}.

These ambitious expansion targets stand against several challenges that the industry is currently facing \cite{bose_questioning_2024}, including cost escalations when building new high-capacity light water reactors (LWR) \cite{rothwell_projected_2022, lovins_us_2022}, uncertainty regarding the availability of so-called "small modular reactors" (SMR) and other "advanced" reactor concepts \cite{steigerwald_uncertainties_2023} as well as a limited actor base \cite{markard_destined_2020}. These challenges raise concerns about whether the nuclear industry will be able to live up to the high expectations raised by above-mentioned announcements.

Currently, several "novel" concepts are being developed. According to a 2023 report by the U.S. "Committee on Laying the Foundations for New and Advanced Nuclear Reactors" \cite[p.43]{committee_on_laying_the_foundation_for_new_and_advanced_nuclear_reactors_in_the_united_states_laying_2023}, these concepts must fulfill four conditions to be competitive in an energy system with a high share of fluctuating renewables: They must be 1) "affordable for owner investment without 'betting the company'", 2) "economically competitive with other technologies [...]", 3) "socially acceptable", and 4) "commercially available." This coincides with required overnight construction costs in the range of 2,000 to 6,000 USD per kW and the establishment and diffusion of non-electrical uses for nuclear power plants \citep{fattahi_analyzing_2022, ingersoll_cost_2020, committee_on_laying_the_foundation_for_new_and_advanced_nuclear_reactors_in_the_united_states_laying_2023}. However, there is a considerable discrepancy between cost assumptions or projections and actual costs of high capacity LWR projects, especially in OECD countries \citep{goke_flexible_2025}. For SMR and non-LWR concepts, limited operational experience hinders the availability of accurate cost and other performance-related data \citep{bose_questioning_2024}. Especially for SMR concepts, manufacturers' cost projections appear somewhat optimistic compared to conventional production theory \citep{steigerwald_uncertainties_2023, kim_challenges_2026}. Nonetheless, smaller sizes, non-electrical uses, and other benefits of nuclear, such as reliability and flexibility, could increase the competitiveness of both SMR and other concepts \citep{buongiorno_future_2018, brown_engineering_2022, black_prospects_2023, carelli_economic_2010}.

Previous studies on the potential viability of nuclear power in decarbonized energy systems limit their assessments to existing nuclear technologies (i.e., high capacity LWRs) and electricity production \cite{jenkins_benefits_2018, bistline_modeling_2023, goke_flexible_2025}, consider only a limited number of low-carbon flexibility options \cite{duan_stylized_2022, koivunen_effect_2024, fattahi_analyzing_2022, soto_dispatch_2022}, limit their analysis to cost parameters \citep{kim_challenges_2026, steigerwald_uncertainties_2023, van_hee_economic_2024} or focus on a single country considered as an energy island \cite{koivunen_effect_2024,duan_stylized_2022,calikoglu_pathway_2023,shirizadeh_low-carbon_2021, satymov_who_2025}. Some literature does refer to "advanced" reactor technologies with links to (thermal) storage options (e.g., \cite{duan_stylized_2022, shirvan_overnight_2022}). However, to our knowledge, a gap exists regarding the detailed assessment of the potential of non-electrical applications for nuclear power in an integrated energy system. Further, a model-based consideration of the potential non-electrical applications of different nuclear reactor concepts is missing due to the limitation of previous studies on a single respective reactor design.

Thus, in this work, we assess the potential of SMR and non-LWR concepts in future energy systems, including non-electrical applications, by following a three-step approach. First, based on an extensive literature assessment, we investigate potential non-electrical applications of nuclear power, e.g., for heat applications, and discuss technology availability. Second, we discuss the potential costs for each technology based on an analysis of 65 references. Third, based on the prior steps, we investigate the economic efficiency of these technologies by integrating them into a European multi-vector energy system model \cite{goke_anymodjl_2021} with the ability to generate both electricity and heat. Assuming non-LWR technologies become commercially available within the next decede, we find that costs must be reduced substantially, i.e., well below current LWR project costs in Europe. These reactor concepts are applied mainly for heat provision for which competitive low-carbon technologies are similarly expensive.

The remainder of this work is structured as follows. Section \ref{background} provides the main results of the literature assessment on use cases and technology availability, and of the cost data analysis. Section \ref{method} introduces the applied energy system model and relevant assumptions and data. The results of the energy system model are presented in Section \ref{results}, and are discussed in Section \ref{discussion}. Section \ref{conclusion} concludes. Further background information is provided in the Appendix.

\section{Background on Nuclear Reactor Technologies} \label{background}

This section gives an overview of potential non-electrical use cases for nuclear power and available reactor concepts. Further, we provide an assessment of cost projections for these currently commercially unavailable reactor concepts.

\subsection{Potential Non-electrical Applications of Nuclear Power}\label{use_cases}

Despite the dominance of LWRs in today's global nuclear fleet, which is mainly used to generate electricity \cite{schneider_world_2024}, the cogeneration of electricity and heat has been the subject of literature and technology development for decades, as in general, nuclear reactors are already "heat generation devices" \cite[p. 23]{csik_nuclear_1997}. \citet{csik_nuclear_1997} further differentiate between reactors intended for heat generation only and those that can also produce electricity, allowing some degree of flexibility. For example, in a recent document, the European Commission stresses the baseload and flexible nature of nuclear power plants that provide "heat to end users in the industry and household sectors [...] today" \citep{ec_commission_2025}[p.22]. However, cogeneration experience remains limited to individual applications; see Appendix \ref{secA2}. This results from the low-level output temperature from LWRs, which limits economic use cases. However, the output temperature could theoretically be raised at the expense of electricity generation \cite{safa_heat_2012}.

\begin{figure}
    \centering
    \includegraphics[width=\linewidth]{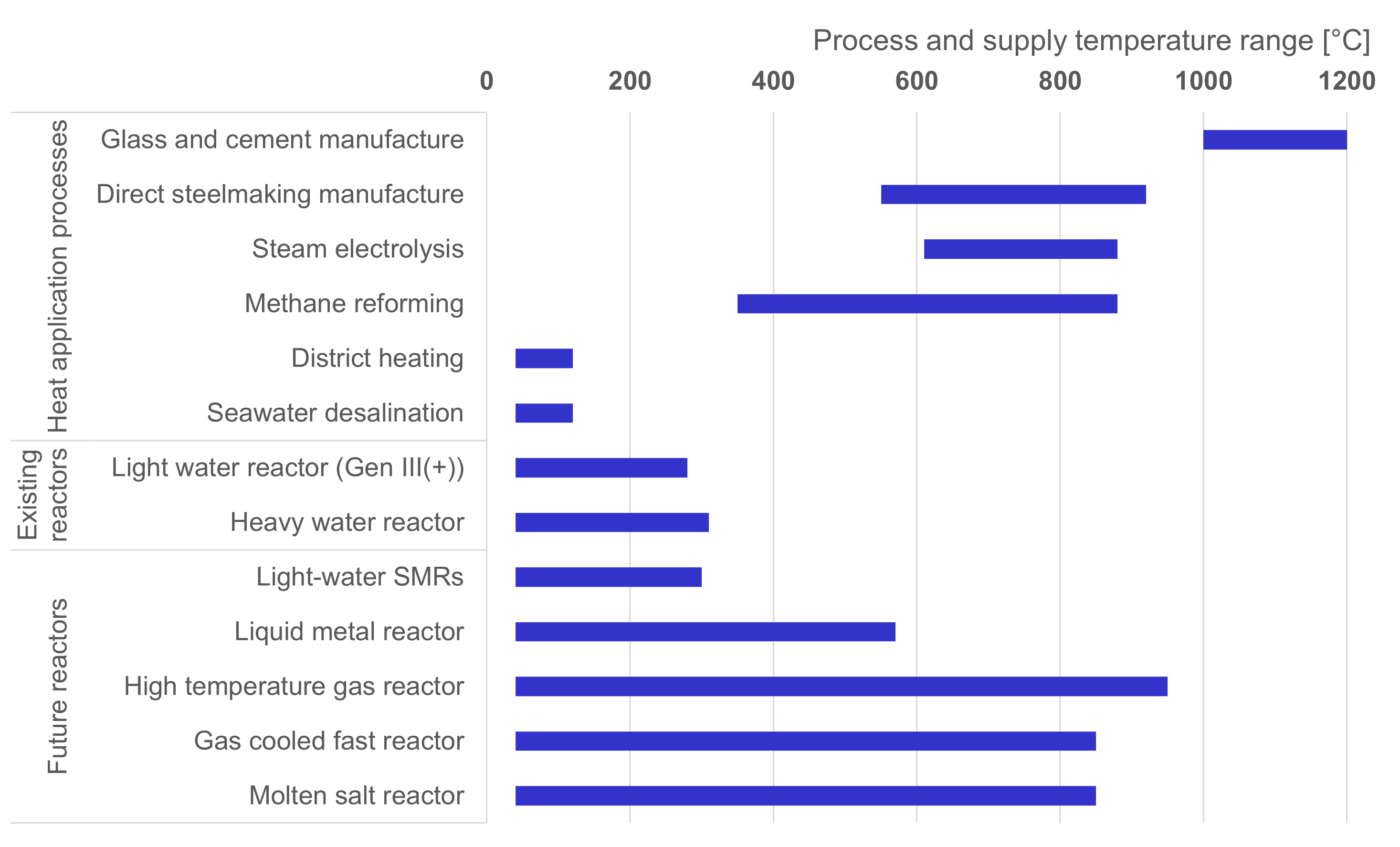}
    \caption{Potential industrial applications for existing and future reactor types, own depiction based on \cite[p. 12]{iaea_opportunities_2017}}
    \label{fig_applications_intro}
\end{figure}

The remainder of this section provides an overview of the potential applications of nuclear power plants for non-electrical uses, such as the provision of industrial and district heat. An overview of possible applications and potential reactor types is shown in Figure \ref{fig_applications_intro}, and several agency reports provide detailed overviews of potential applications of nuclear cogeneration \cite{iaea_market_2002, iaea_opportunities_2017, iaea_guidance_2019, oecd_beyond_2022}. 
A list of nuclear power plants used for non-electrical energy provision is provided in Appendix \ref{secA2}, showing that historical applications are limited to individual plants.

\subsubsection{District Heat}\label{district_heat}

Nuclear power reactors have been supplying low-temperature heat up to 100°C for district heating for decades \cite{nilsson_secure_1978, leppanen_review_2019}. While there are issues related to acceptance, e.g., addressing the proximity to populated areas \cite{committee_on_laying_the_foundation_for_new_and_advanced_nuclear_reactors_in_the_united_states_laying_2023}, regulation, e.g., regarding the establishment of emergency planning zones \cite{leppanen_review_2019, black_prospects_2023}, as well as technology, e.g., the challenge of efficiency losses when transferring heat over long distances \cite{safa_heat_2012, brown_engineering_2022}, SMRs and non-LWR concepts are thought to be especially applicable to this use case \cite{leppanen_review_2019, satymov_who_2025}. There have been several attempts to design reactors specifically for district heating in Sweden, Finland, and China \cite{nilsson_secure_1978, leppanen_review_2019}. Mostly in Eastern Europe, there exists decade-long experience with district heating via nuclear cogeneration, for example, at the Kozloduy plant in Bulgaria, the Bohunice V-2 plant in Slovakia, or the Paks plant in Hungary. The heat is used for small demand centers nearby \cite{kupitz_small_2001, oecd_beyond_2022}. Additionally, in 2019, two Westinghouse AP1000 reactors began providing district heat to the city of Haiyang in China \cite{kraev_city_2021}.

\subsubsection{Process Heat for Industry}\label{process_heat_industry}

Decarbonizing the provision of heat for industrial processes such as cement production and steel manufacturing is challenging as most processes require high temperature levels of several hundred °C and a near-constant provision of energy \cite{pisciotta_current_2022}. For decades, it has been argued that nuclear power can provide both the required heat and capacity. For example, current high-capacity LWRs can theoretically produce heat ranging from 200 to 300°C, applicable for pulp, paper, and textiles manufacturing or seawater desalination (see Section \ref{other}), and advanced gas-cooled reactors currently operating in the U.K., can provide heat up to 650°C, a temperature level relevant for the chemical industry \cite{csik_nuclear_1997}. So far, however, experience is limited to individual, mostly low-capacity demonstration projects that use mainly excess heat from electricity generation. Examples include the Calder Hall gas-cooled reactors, built in the 1950s in the U.K., which provided steam to the nearby Windscale reprocessing plant, or the Gösgen plant in Switzerland, which has been supplying heat to a nearby cardboard factory using 1\% of its steam since 1979 \cite{iaea_industrial_2017}. In Canada, the Bruce heavy-water nuclear plant provided heat for its own usage but also to a greenhouse, an ethanol plant, and a plastic film production plant, amongst others, covering distances of over six kilometers \cite{kupitz_small_2001}. The Norwegian boiling heavy water research reactor in Halden provided heat that would have otherwise been dissipated into a nearby river to a paper mill until the reactor was closed in 2018 \cite{iaea_guidance_2019}.

\subsubsection{Hydrogen Production}\label{hydrogen}

The use of hydrogen in industrial processes, but also for energy storage, is becoming increasingly relevant in decarbonization efforts \cite{oecd_role_2022, pisciotta_current_2022}. In general, there exist two ways in which nuclear energy could be utilized for hydrogen production. Either via heat provision for steam methane reforming \cite{bartels_economic_2010} and for other thermo-chemical processes such as sulfur-iodine hydrogen production at temperatures greater than 750°C \cite{nuttall_fossil_2020, doctor_evaluation_2006}, or via the generation of electricity for an electrolyzer, which would lead to further competition with renewables and raise efficiency concerns \cite{nuttall_fossil_2020, pregger_prospects_2009}.

\citet{pinsky_comparative_2020} assess the applicability of so-called nuclear hybrid systems for hydrogen production and conclude that most technically feasible approaches for which nuclear could become competitive remain commercially unavailable. This assessment is supported by the lack of commercial nuclear hydrogen production projects, such as the scrapped U.S. Department of Energy project \cite{bartels_economic_2010, schneider_world_2023}.

\subsubsection{Other Applications}\label{other}

Other non-electrical nuclear power applications include marine propulsion technology, most recently applied in Russia for the Akademik Lomonosov \cite{schneider_world_2024}, and in nuclear submarines \cite{rawool-sullivan_technical_2002}. Other nuclear propulsion devices are used in outer space \cite{iaea_market_2002}. Seawater desalination is a particular use case for nuclear power applied at several locations \citep{al-othman_nuclear_2019}. Different desalination processes require different forms of energy, such as electricity (reverse osmosis) or heat (multi-effect distillation or multi-stage flash) \cite{ingersoll_extending_2014}, and are thus applicable to different reactor types. Required temperatures range from 200 to 300 °C \cite{csik_nuclear_1997}. Operational experience is limited to Japan, where since 1978, one BWR and seven PWRs use district heat for small seawater desalination plants, to Kazakhstan, where the first nuclear desalination facility at Aktau operated a sodium-cooled fast reactor from 1973 to 1999, to several Russian civilian nuclear-powered ships, operated since the 1980s, and to the Diablo Canyon power plant in California \cite{faibish_application_2002, kupitz_small_2001}. 

\subsection{Nuclear Technologies and Availability}\label{availability}

The nuclear industry has been in decline for several decades, especially in OECD countries, as only a few projects were realized, which were behind schedule, well over budget, and resulted in a decline of major industry actors (e.g., the bankruptcy of Westinghouse) \cite{markard_destined_2020}. Most discussed nuclear technologies are either unavailable or economically disadvantageous compared to other low-carbon technologies, and thus, new-builds are mainly realized in state-owned spheres, especially in China and Russia \cite{lovins_us_2022, schneider_world_2024}. Regardless, ambitious expansion plans for nuclear power are being publicly announced \cite{wnn_ministerial_2023}, and nuclear power continues to play a major role in energy scenarios \cite{byers_ar6_2022}. While the global nuclear fleet is dominated by high-capacity LWRs, electrical and non-electrical applications are envisioned to be implemented by using LWRs with capacity of less than 300 MW\textsubscript{el} (referred to here as "small-modular reactors" (SMRs)) and so-called "advanced" or "Generation IV" reactors, that comprise several non-LWR concepts \cite{bose_questioning_2024, oecd_beyond_2022}. These alternative concepts must also fulfill four prerequisites to be viable for large-scale deployment in a future decarbonized energy system. They must be 1) affordable, 2) economically competitive with other low-carbon technologies, 3) socially acceptable, and 4) available on a commercial scale \cite{committee_on_laying_the_foundation_for_new_and_advanced_nuclear_reactors_in_the_united_states_laying_2023}. In the following, we briefly discuss these prerequisites for often discussed reactor types, while omitting questions of social acceptance. See Appendix \ref{secA1} for more technical details and information on past and ongoing research and development for the here-discussed non-LWR technologies.

\subsubsection{Light Water Reactors}

\paragraph{High-capacity LWR}\label{lwr}

High-capacity light water reactors (LWR), i.e., reactors exceeding capacities of 300~MW\textsubscript{el}, referred to also as "Generation III(+)" reactors, dominate global reactor fleets today \cite{schneider_world_2024}. These reactors are usually operated as "baseload" plants with relatively high capacity factors and are mostly limited to electricity production. As discussed in Section \ref{use_cases}, individual plants have also produced district or process heat. Currently, the construction of new plants is limited mostly to Russian and Chinese projects and will have to compensate for aging reactor fleets facing closure \cite{schneider_world_2024}. Projects in Europe and the U.S. are pained by escalating costs and construction times \cite{rothwell_projected_2022}, and a limited actor base shows an industry in decline \cite{markard_destined_2020}. Thus, it is highly questionable whether these reactor types will be available on a sufficient scale for energy system decarbonization in the coming decades \cite{bose_questioning_2024}. Independent and reliable cost data is available for Western (OECD) countries \citep{buongiorno_future_2018}. At current cost levels, LWRs are not economically competitive against renewable energies in a low-carbon energy system \citep{price_role_2023, goke_flexible_2025}. Nonetheless, assuming that aging fleets were to be replaced with new reactors, flexible operation and heat provision are technically feasible and could bring benefits to a system characterized by a high degree of renewables, such as lower system operational costs, increased reactor owner revenues, and reduced curtailment of renewables \cite{jenkins_benefits_2018}.

\paragraph{Small Modular Reactors (Low-capacity LWRs)}\label{SMR}

SMRs are envisioned to reduce costs compared to LWRs mainly due to modularization and factory production of components, facilitated regulation, passive safety systems, and shorter construction times that would reduce capital costs \cite{robb_stewart_construction_2023, steigerwald_uncertainties_2023, kim_challenges_2026}. Further, SMRs are proposed for niche applications, such as energy-intensive operations in remote locations \cite{black_small_2021, sanongboon_techno-economic_2024, al-salhabi_feasibility_2024} or microgrids \citep{testoni_review_2021, michaelson_review_2021}. In 2014, \citet{liu_technology_2014} assumed that light water SMRs could be on a high technology readiness level of 7 to 8 and could be elevated to level 9, i.e., the operation of a functioning system, within the next ten years because of experience of similar reactor concepts in the 1950s and 1960s. However, the transferability of knowledge from early concepts should be questioned, primarily due to increased safety requirements for nuclear reactors \citep{ramana_small_2021}. Operation and construction experience of SMRs is limited to a small number of research and prototype reactors that have, like their high-capacity counterparts, also suffered from decade-long construction times or are still in the licensing stage and remain years away from actual operation \cite{bose_questioning_2024, pistner_sicherheitstechnische_2021}. Additionally, to be economically competitive with other low-carbon technologies, SMRs would first have to overcome diseconomies of scale, and even then, cost projections by manufacturers are likely overly optimistic \cite{steigerwald_uncertainties_2023}. Nonetheless, SMRs are considered for potential non-electrical applications. For example, the NuScale VOYGR is expected to be able to provide a pre-heated feedstock of heat for refinery processes at around 300 °C \cite{ingersoll_extending_2014}, and it is supposed to be compatible with high-temperature electrolyzers for hydrogen production \cite{ingersoll_extending_2014, henderson_us_2006, bartels_economic_2010}. The only VOYGR order, however, was canceled in late 2023 due to cost escalations \cite{gardner_nuscale_2023}. Other projects, such as the Argentinian CAREM-14 reactor, have been scrapped, or are facing increasing delays, like the Ontario Power Generation's BWRX-300 project in Canada \citep{schneider_world_2025}. 

\subsubsection{Reactors with Fast Neutron Spectra}\label{fbr}

Reactors operating on a fast neutron spectrum, so-called fast breeders (FBR), were historically considered to be part of historical nuclear capacity expansions in a so-called "plutonium economy" \cite{seaborg_plutonium_1970, midttun_negotiating_1986}. The envisioned fuel cycle would reuse plutonium from spent fuel via reprocessing and MOX-fuel fabrication to decrease the demand for natural uranium (referred to as the closed fuel cycle) \cite{bunn_economics_2003}. Proponents envisaged that FBRs would replace LWRs and energy would become "too cheap to meter" \citep{strauss_remarks_1954}[p. 9]. There have been several attempts to establish FBRs for large-scale energy generation, but none have succeeded on a commercial scale \cite {cochran_fast_2010}. The following section briefly addresses the two most common types, sodium-cooled fast reactors (SFR) and lead-cooled fast reactors (LFR). Gas-cooled fast reactors are omitted due to the lack of prototypes, see Appendix \ref{secA1}.

\paragraph{Sodium-cooled Fast Reactors}

SFRs were amongst the first reactor concepts to be developed in the wake of feared uranium source depletion that would ultimately render LWRs inoperable. However, while projected nuclear capacity expansions did not formulate and new uranium deposits were discovered, the economic deficiencies and the high technological complexity of SFRs remained, and a large-scale commercial deployment never happened \cite{schulenberg_fourth_2022, ohshima_sodium-cooled_2016}. SFR projects are characterized by project failures rather than operational reactors, and due to, e.g., the high reactivity of sodium with air and water, operational reactors are unreliable \cite{bose_questioning_2024}. Regardless, four SFRs are currently under construction \cite{iaea_nuclear_2022}, and increased thermal efficiency coupled with outlet temperatures of 400 to 450°C could make SFRs applicable for non-electrical applications in niche markets, despite economic deficiencies stemming mostly from additional safety requirements \cite{pistner_analysis_2024}. Current projects include a 345 MW\textsubscript{el} SFR coupled with a molten salt tank, designed by U.S.-company TerraPower, currently undergoing pre-licensing activities with regulators \cite{schneider_world_2024}. 

\paragraph{Lead-cooled fast reactors}

Lead-cooled fast reactors (LFR) operate similarly to SFRs, but use lead as a coolant. Lead is less reactive than sodium and could thus reduce operational risks and necessary safety barriers \cite{alemberti_lead_2021, soto_dispatch_2022}. However, operational experience is almost exclusively limited to Russia, where LFRs were experimented with for submarine propulsion, and a single reactor is currently under construction \cite{schulenberg_fourth_2022}. Further, a demonstration LFR with an accelerator-driven system is currently under development in Belgium (MYRRHA) \cite{pistner_analysis_2024}. Due to limited operational experience, cost information is scarce. However, it is expected that LFRs could be cheaper than SFRs due to reduced safety risks, and if a closed fuel cycle can be implemented, LFRs might also compete with current LWRs \cite{alemberti_lead_2021}.

\subsubsection{High temperature reactors}\label{htr}

Like other non-light water reactor technologies, operational experience for high-temperature reactors (HTR) is also limited. Developed in the 1960s to operate on thermal spectra, different reactor concepts, such as the Dragon reactor in the UK or the pebble-bed reactor THTR-300 in Germany, were constructed as prototypes but faced significant technological challenges and were thus abandoned \cite{pistner_analysis_2024}. Current operational reactors are limited to China, where two prototypes based on the German pebble-bed design are running. In the U.S., X-energy is working on an 80 MW\textsubscript{el} HTR, the Xe-100, which is still in the design phase \cite{zhang_shandong_2016, bose_questioning_2024}. Given their high operational temperature, HTRs are envisioned to be applicable for high-temperature heat provision of up to 1000°C \citep{buongiorno_future_2018, pistner_analysis_2024}. For example, a proposed design for a gas turbine modular helium reactor could produce outlet temperatures of 950°C and could be applied for sulfur-iodine hydrogen production with a process efficiency as high as 51\% \cite{schultz_large-scale_2003}. Potential cost disadvantages compared to LWRs, stemming from more complex designs, could, theoretically be offset by the provision of valuable high-temperature heat \cite{stewart_capital_2021}.

\subsubsection{Molten salt reactors}\label{MSR}

Molten salt reactors (MSR) use molten salt as a coolant \cite{riley_molten_2019}. MSRs were initially developed for aircraft propulsion, and concepts vary regarding moderator, fuel, neutron spectrum, and fissile material. There are no commercially operational MSRs, but several concepts are currently under development in China, Switzerland, and Russia, amongst others \cite{pistner_analysis_2024, hejazi_small_2025}. MSRs could become applicable for heat provision in decarbonized energy systems as the proposed salts melt at around 500°C and boil at around 1400°C, allowing for outlet temperatures ranging from 600 to 700°C \cite{riley_molten_2019, pistner_analysis_2024}. Most MSR concepts are still in the early development stages \citep{hejazi_small_2025}, so information on their economic competitiveness is limited. While in the 1970s, capital costs of MSR were estimated in the same range as LWRs \cite{rosenthal_molten-salt_1970}, current concepts are estimated to be cheaper. For example, the so-called "Advanced High-Temperature Reactor," also referred to as "Fluoride High-Temperature Reactor," is estimated to cost less than 1000 USD per kW, and the ThorCon concept MSR, developed by company Martingale, is envisioned to produce electricity for as little as 30 to 50 USD per MWh \cite{wna_molten_2021, pistner_analysis_2024}.

\section{Methodology}\label{method}

The following section describes the methodology used to assess the potential of non-LWR concepts in a future European energy system. First, we introduce the applied energy system model. Then, we motivate the selection of four reactor technologies for our application based on previous sections, and finally, we describe our cost assumptions based on our vast cost analysis.

\subsection{Applied Energy System Model}\label{method_model}

\begin{figure}
\centering
\includegraphics[width=\linewidth]{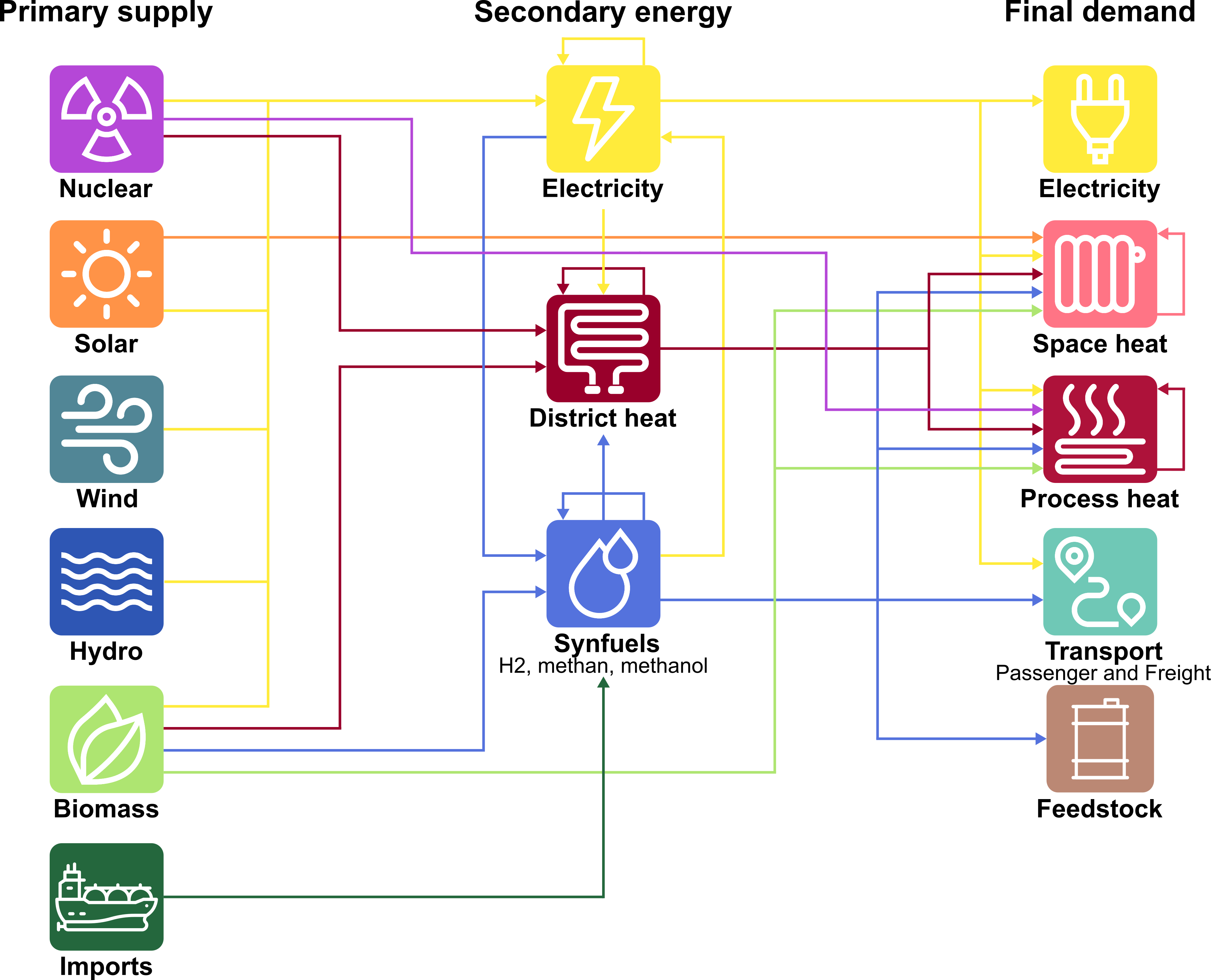}
\caption{Energy vectors considered in the applied model.}
\label{fig:overMod}
\end{figure}

This work applies a comprehensive energy system model to assess different reactor concepts \citep{goke_liquid_2025}. It utilizes the AnyMOD.jl framework for energy system modeling \cite{goke_anymodjl_2021}. Fig. \ref{fig:overMod} gives an overview of the energy vectors in the model. Sources of primary energy supply in the left column include nuclear and renewables, i.e., solar, wind, hydro, and biomass. The middle column of Fig. \ref{fig:overMod} lists the secondary energy carriers that can be generated from the primary supply and include electricity, district heat, and various synthetic fuels, including hydrogen, synthetic methane, and methanol. Finally, the right column shows the types of final energy demand that the model must cover. These include final demands for electricity, space heat, and process heat, respectively. The demand for process heat is further subdivided into three levels: low temperature up to 100°C, medium temperature from 100°C to 500°C, and high temperature above 500°C. In addition, the model also covers the demand for transport services, specifically private road, public road, and public rail transport for passengers and freight transport on heavy road, light road, and rail. Finally, it includes an exogenous fuel demand covering aviation, navigation, and feedstocks in the industry. As indicated in the figure, the model considers nuclear power not just as a source of electricity but also for the provision of district and process heat, as described in Section \ref{method_technologies}.

The model is formulated as a linear optimization problem that decides on capacity investment and operation of capacities to satisfy a fixed final demand at the lowest cost possible. Investment options are different technologies to transport, convert, store, and generate different energy carriers, see Fig. \ref{fig:overMod}. Technology data and demand are based on projections for the year 2040. Appendix \ref{secA4} gives an overview of the considered technologies besides nuclear.

The model pursues a stochastic approach to capture the effect of uncertain weather conditions on renewable generation and demand based on \citet{goke_liquid_2025}. Weather foresight is limited to a month, and the model explicitly models 16 different months to account for extreme seasonal variations. However, a robust formulation ensures reliability for all possible combinations of these months, resulting in 12 different climate years. The model employs time-series clustering to reduce the underlying time series from 8760 to 2856 hours, while maintaining daily patterns.

The spatial scope covers the European Union (excluding isolated island states), the U.K., Switzerland, Norway, and the Balkan region. The temporal scope consists of a single future year in a greenfield setting. Due to their long lifetimes, existing hydropower plants and power grids are available without additional investment.

The spatial resolution of the model varies by context. For the characteristics of renewables, all types of heat and hydrogen, the spatial scope is split into 96 clusters. For the electricity balance and transport demand, the scope is split according to the zonal configuration of the European power market. Appendix \ref{secA5} details the spatial configuration and the available transport infrastructure to exchange energy between regions. \citep{goke_flexible_2025, goke_how_2023}

\subsection{Applied Nuclear Technologies and Technical Assumptions}\label{method_technologies}

From the analysis of use cases and available technologies (see Sections \ref{use_cases} and \ref{availability}) and the available cost data presented in Section \ref{method_literature} below, we derive the necessary input parameters for a total of four different reactor technologies to be implemented into the model. These are high-capacity (advanced) light-water reactors (LWR), low-capacity light-water reactors (SMR), sodium-cooled fast reactors (SFR), and high-temperature gas-cooled reactors (HTR). Each technology is assumed to be able to supply a particular form of heat in addition to electricity, see Table \ref{tab:input_data}.


Nuclear power is sometimes portrayed as a dispatchable technology that could bring benefits to an energy system characterized by a high penetration from renewables \cite{jenkins_benefits_2018, lynch_nuclear_2022, aunedi_system-driven_2023}.  Further assumptions relate to individual reactor technologies' abilities to switch operational modes. We assume that reactors can flexibly switch from electricity to heat provision and vice versa without any downtime or incurred cost. We allow all plants apart from high-capacity LWRs to operate in three distinct modes that influence the thermal efficiency of the reactors. A full heat mode increases thermal efficiency by an average of 26\% compared to the mode with the highest output of electricity. A medium mode increases thermal efficiencies by 16\%. LWRs may operate only on the medium and low heat modes. The reactors in the model operate without ramp-up time at a universal 90\% capacity factor independent of the chosen mode.

\subsection{Nuclear Cost Data}

\subsubsection{Cost Data Analysis} \label{cost_analysis}

Over the last decade, nuclear power has gained momentum in energy models as a low-carbon technology \citep{bose_questioning_2024}. Many of these models assume substantial cost reductions for nuclear in the coming decades down to overnight costs of 5,000~or even 2,000~USD per kW (e.g., \citet{bistline_modeling_2023}) and highlight the necessity of cost reductions for nuclear to be competitive in energy systems with high shares of renewables \cite{duan_stylized_2022, baik_californias_2022}.

Thus, to assess the economic efficiency of non-light water reactors in the energy system model (see Section \ref{method_model}), it is first necessary to determine relevant cost data and other relevant information on potentially viable reactor types. Consequently, we conducted a literature analysis of cost projections and assumptions for various reactor types. Previous research has shown that for high-capacity LWRs, cost projections and experience differ strongly \citep{goke_flexible_2025}. For SMR concepts, \citet{steigerwald_uncertainties_2023} and \citet{kim_challenges_2026} show that to-be-expected costs and industry expectations also vary substantially. This is in line with findings by \citet{abdulla_expert_2013} who determine that cost projections for nuclear are frequently overly optimistic and diverge strongly.

The here presented cost analysis is based on 65 references  \citep{rodriguez_astrid_2018,rothwell_projected_2022,budi_fuel_2019,green_smr_2019,iaea_advances_2020,stein_advancing_2022,dixon_advanced_2017,schultz_large-scale_2003,nian_economic_2020,faibish_status_2019,iaea_current_2001,birch_ebr-ii_2019,hibbs_future_2018,holdman_small_2011,samalova_comparative_2017, brooking_design_2015, wang_micro-reactor_2019,tass_cost_2021,lu_fully_2019,us_doe_pathways_2023,short_deployability_2018,stewart_capital_2022,wnn_nuscale_2020,schlissel_eye-popping_2023, foss_cost_2021,us_doe_capital_2020,smr_start_economics_2021,stewart_pathways_2020,gandrik_assessment_2012,buongiorno_future_2018,stewart_economic_2021,us_doe_modular_1993,cleantech_catalyst_ltd_eti_2018,ganda_economic_2015,ganda_report_2018,david_e_shropshire_advanced_2009,boardman_economic_2001,hoffmann_estimated_2004,engel_conceptual_1980,holcomb_advanced_2011,iaea_comparaitve_2023,schroders_energy_2018,van_heek_incogen_1996,tolley_economic_2004,stewart_capital_2021,ingersoll_cost_2020,karakosta_renewable_2013,asuega_techno-economic_2023,ec_energy_2014,moir_cost_2002,pistner_analysis_2024,zandt_small_2024,balanin_assessing_2024,khan_technological_2025,lovering_techno-economic_2023,testoni_review_2021,abdulla_expert_2013,kellison_economics_2025,richards_economic_2017,noland_overview_2025,li_investigation_2019,soto_dispatch_2022,alonso_economic_2025,alonso_process_2014,van_hee_economic_2024}. About 51\% are so-called grey literature, mostly industry and research organization reports. Peer-reviewed literature, i.e., journal and conference papers, account for about a third. Almost 45\% of the references are only five years old. A detailed overview of cost data and the corresponding references is given in Appendix \ref{secA3}.

Overnight construction cost (OCC) data are provided on different "readiness levels," i.e., either as so-called "first-of-a-kind" (FOAK) or "n-th-of-a-kind" (NOAK) estimates. FOAK cost estimates acknowledge that the first built reactor of a given type will likely experience unexpected complications and thus be more expensive than subsequent projects of the same design. NOAK cost estimates consequently assume that learning processes reduce unit costs \cite{lloyd_expanding_2020}. Learning rates for nuclear vary and, depending on the assessment and data, range from 2 to 15\% \cite{stewart_capital_2022}. However, negative learning rates have also been observed \cite{grubler_costs_2010}. \textit{n} indicates the number of units produced to reach a certain cost level. In general, NOAK costs are calculated following Eq. \ref{eq_noak} with the cost for the n-th produced unit \textit{NOAK\textsubscript{n}}, the cost for the first produced unit \textit{FOAK}, learning rate \textit{x} and doubling \textit{d} of production volume \textit{n} \cite{lloyd_expanding_2020}.

\begin{equation} \label{eq_noak}
\begin{gathered}
NOAK_n=FOAK*(1-x)^d \\
n = 2^d
\end{gathered}
\end{equation}

Regardless of the readiness level, every value constitutes either an estimation or projection, as none of the discussed reactors have been built. Furthermore, from our gathered literature, we find that only \citet{stewart_economic_2021} and \citet{stewart_capital_2022} attribute a concrete value to \textit{n}, i.e., 10. \citet{noland_overview_2025} investigate the required number of built reactors for four types of SMRs. They find that while only eight units of the Rolls-Royce SMR (470~MWe) would suffice to reach announced cost projections, the same approach would require 41,025 VOYGR units. Additionally, in all three references, the FOAK starting point is far off from historical precedence, especially for high-capacity LWRs \cite{goke_flexible_2025}. Regarding learning rates and the feasibility of industry cost reduction projections, refer to \citet{steigerwald_uncertainties_2023} who assess learning rates and potential cost reductions for 19 SMR and advanced reactor concepts. For the following, cost data was normalized and adjusted to 2019-USD following the "cost escalation approach" by \citet{abou-jaoude_literature_2023} to ensure comparability. \footnote{If not otherwise indicated, all USD values are in 2019-USD.}

\begin{figure}
    \centering
    \includegraphics[width=\linewidth]{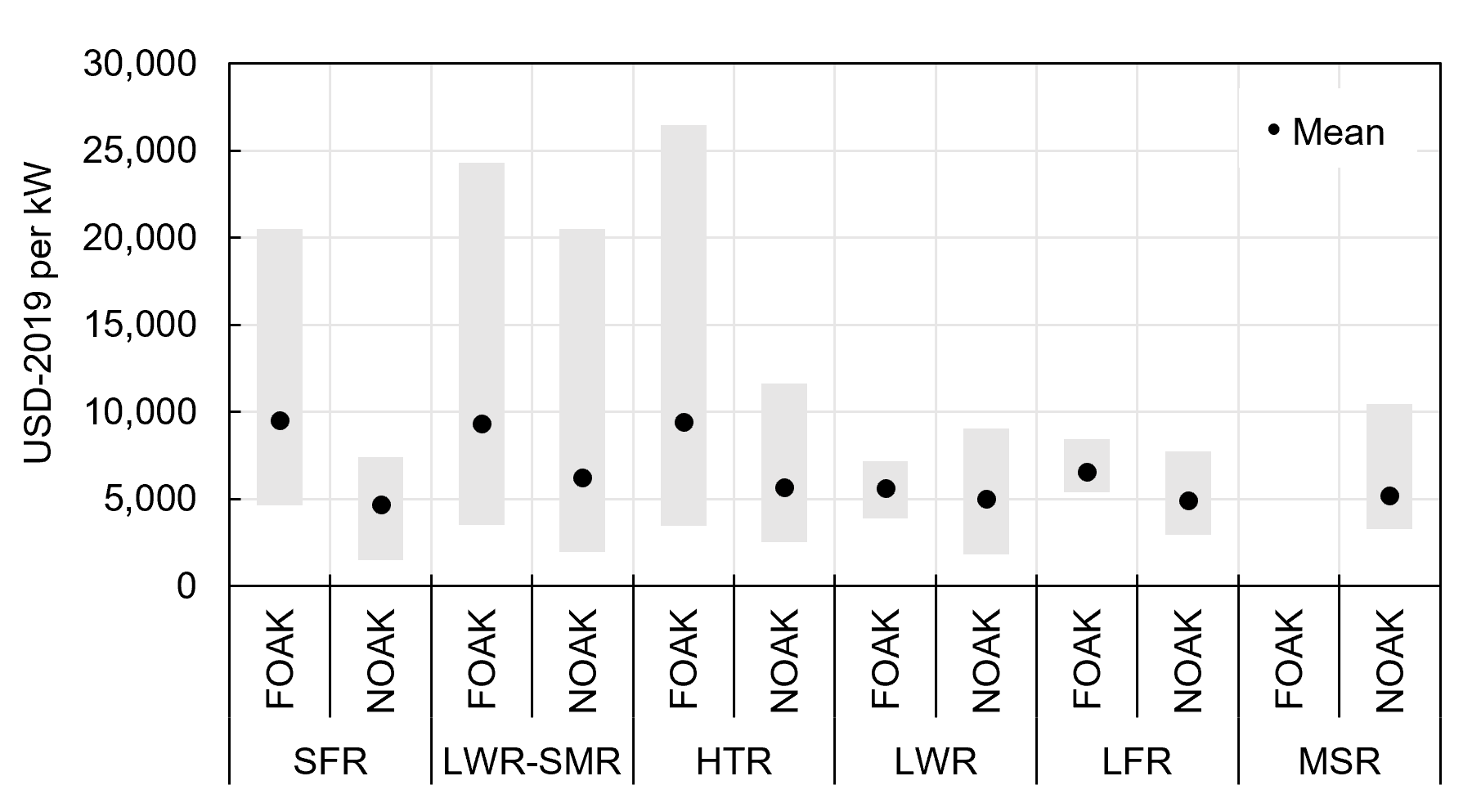}
    \caption{Distribution of overnight construction cost data in USD-2019 per kW for different reactor types. Note that the single data point for the molten salt fast reactor provided by \citet{brooking_design_2015} is not shown here. Abbreviations used: FOAK = "first of a kind"; NOAK = "n-th-of-a-kind"; SFR = sodium-cooled fast reactor; LWR = high-capacity light water reactors; LWR-SMR = low-capacity light water reactor (aka small modular reactor); HTR = high temperature reactor; LFR = lead-cooled fast reactor; MSR = molten-salt reactor.}
    \label{fig_cost_occ_boxes}
\end{figure}

\begin{figure}
    \centering
    \includegraphics[width=\linewidth]{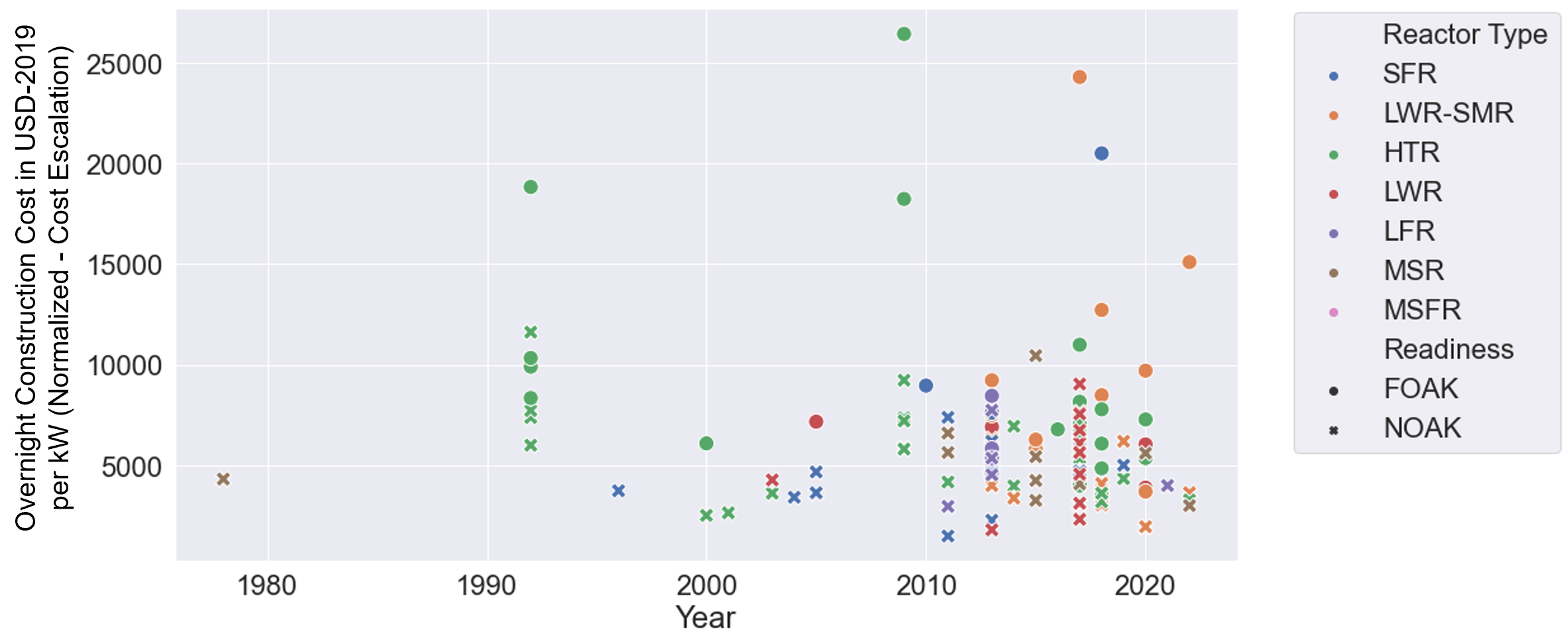}
    \caption{Distribution of overnight construction cost data in USD-2019 per kW for different reactor types and readiness levels over the reference years in literature. If no information is provided, the publication year was taken. Abbreviations used: FOAK = "first of a kind"; NOAK = "n-th-of-a-kind"; SFR = sodium-cooled fast reactor; LWR = high-capacity light water reactors; SMR = small modular reactor; HTR = high temperature reactor; LFR = lead-cooled fast reactor; MSR = molten-salt reactor; MSFR = molten-salt fast reactor.}
    \label{fig_cost_occ_points}
\end{figure}

Figure \ref{fig_cost_occ_boxes} shows the variation of cost data between reactor types and readiness levels. FOAK cost data are not available for MSRs. In general, NOAK estimates are more numerous than FOAK estimates. Further, on average, NOAK estimates are lower than FOAK estimates by several thousand USD per kW. This indicates widespread expectations of future cost digressions despite lack of precedence \cite{koomey_reply_2017, grubler_costs_2010, markard_destined_2020}. Figure \ref{fig_cost_occ_points} shows the distribution of OCC values over time. The x-axis shows either the publication year or, if available, the stated reference year. Tables \ref{tab:occ_data}, \ref{tab:om_fix_data} and \ref{tab:om_var_data} in Appendix \ref{secA3} summarize cost data for OCC, as well as for fixed and variable operational costs for various reactor types. Fewer data points were identified for these cost data than for OCC, and information could not be found for each reactor type.

\subsubsection{Nuclear Cost Assumptions}\label{method_literature}

From obtained OCC data, total capital costs are calculated following \citet{rothwell_economics_2016} by accounting for a uniform seven years of construction time and weighted average cost of capital of 10\%. We conduct a step-by-step scenario approach to determine cut-in costs for different nuclear technologies. In ten deterministic scenarios, we reduce OCC for each considered nuclear technology from the mean FOAK value to the minimum NOAK via linear interpolation. Further relevant input parameters are given in Table \ref{tab:input_data} and reflect mean NOAK values from the literature. Due to the heterogeneity of provided data, OM costs are combined from fixed and variable values following the approach used in \citet{goke_flexible_2025}. These OM costs explicitly exclude fuel costs that are included as a separate value. Further, we assume uniform capacity factors of 90\% and operational lifetimes of 40 years. To match other model input cost parameters, cost data for nuclear given in USD-2019 were converted to EUR-2020 with an inflation rate of 1.4\% and an exchange rate of 0.8929.


\begin{table}[h!]
\begin{tabular}{lllll}
\hline
\textbf{Reactor Technology} & \textbf{OCC$^a$} & \textbf{O\&M cost$^b$} & \textbf{Fuel Cost$^b$}  & \textbf{Energy Carriers}  \\ \hline
LWR                         & 1747 - 5511  & 18.51              & \multirow{4}{*}{8.29} & Electricity \& District Heat \\
SMR                         & 1902 - 9126  & 18.85              &                      & Electricity \& Low-temperature process heat \\
SFR                         & 1446 - 9323  & 18.05              &                      & Electricity \& Medium-temperature process heat \\
HTR                         & 2451 - 9252  & 30.47              &                      & Electricity \& High-temperature process heat \\ \hline
\end{tabular}
\caption{Cost-related input parameters used for each reactor type. $^a$Given in USD per kW. $^b$Given in USD per MWh. Abbreviations used: LWR = High-capacity light water reactor; SMR = Small Modular Reactor (limited to light water technology); SFR = sodium-cooled fast reactor; HTR = High-temperature reactor}
\label{tab:input_data}
\end{table}

\section{Results}\label{results}

In the following section, we first present the results of the model analysis for nuclear capacity expansions before discussing the impact of nuclear expansion on the whole energy system.

\subsection{Overview}

\begin{figure}[h!]
\centering
\includegraphics[width=\linewidth]{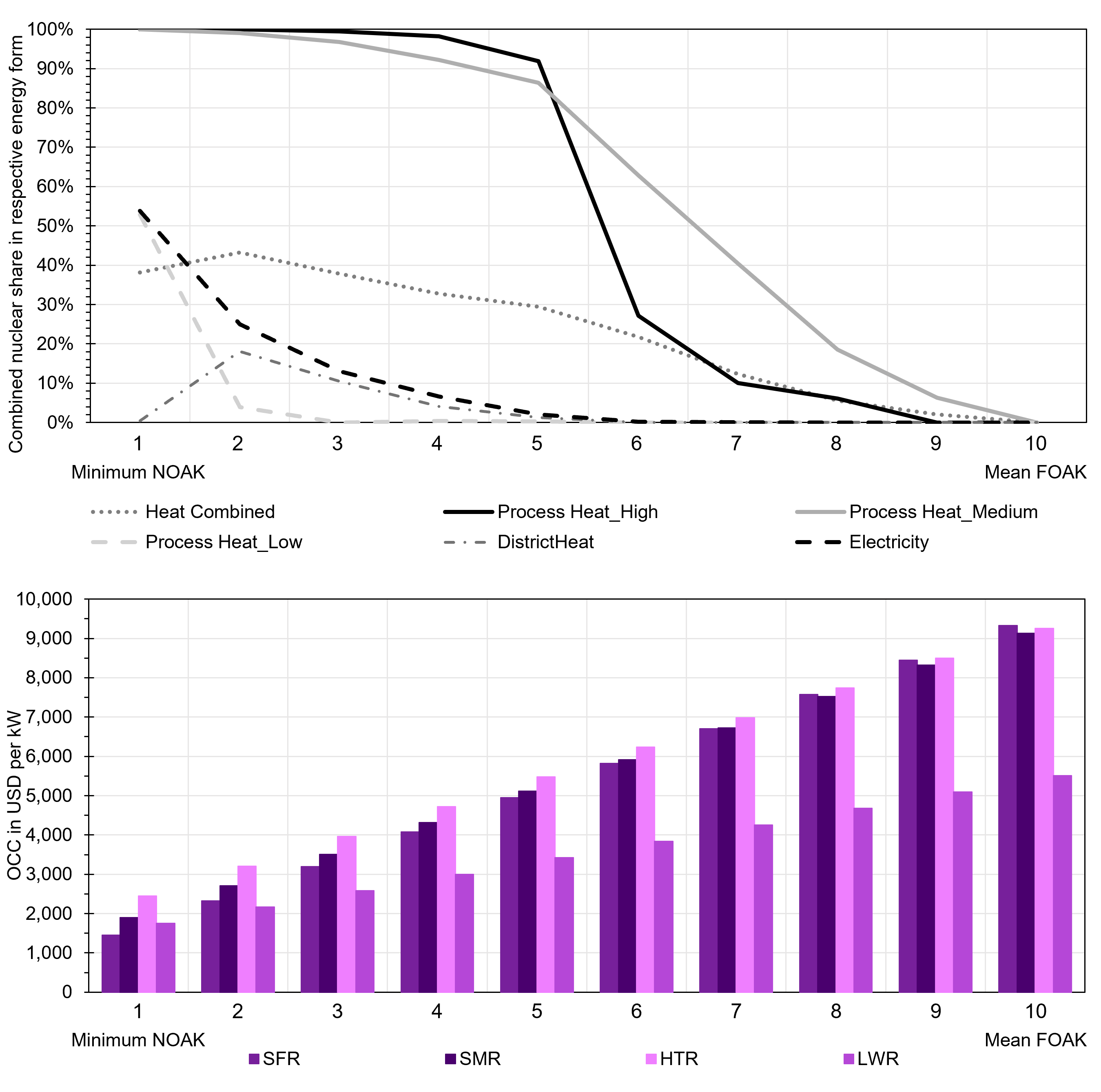}
\caption{Shares of nuclear technologies in electricity and heat provision plotted against assumed OCC values for ten cost scenarios.}
\label{fig:globalResults}
\end{figure}

Overall, we find that with decreasing costs for nuclear, its shares in both electricity and heat provision increase. Figure \ref{fig:globalResults} shows the share of nuclear, i.e., the combined production of all four technologies, for electricity, the combined provision of heat, and all four respective types of heat. It also shows the corresponding stepwise digression of OCC values for each technology.

The share of electricity production from nuclear reaches the current levels in the EU (22.6\% as of 2023 \citep{schneider_world_2024}) once OCC values are almost at the NOAK level (No. 2 in Figure \ref{fig:globalResults}). In this scenario, OCC values are 2,165~USD/kW for LWR, 3,207~USD/kW for HTR, 2,704~USD/kW for SMR, and 2,322~USD/kW for SFR.

In contrast, the nuclear share of heat generation rises continuously and reaches its maximum level of 43.2\%, again when costs are almost at the NOAK level. High and medium process heat, provided by HTRs and SFRs, respectively, reach high shares at higher costs (No. 6 \& 7).

The share of nuclear in low temperature process heat provision remains marginal until costs for SMR fall below 2,704 USD/kW, when the share is determined at 3.8\% (No. 2). The highest share is 52.9\% at the lowest costs of 1,902 USD/kW (No. 1).

It is noteworthy that the nuclear heat share falls between the cheapest (No.1) and second-to-last cost scenario (No. 2). This is driven by the substantial reduction of district heat provision from LWRs from a maximum of 18\% to about 0.43\% in the lowest cost scenario. As shown in Figure \ref{fig:globalResults}, the OCC for LWRs supersede those of SFRs in the minimum NOAK scenario (No. 1). Consequently, the cost-optimizing model selects the superior technology, SFR, to provide the electricity that was previously provided by LWRs. Given the relatively low share of district heat provision by LWRs in other scenarios, and the model's switch to another reactor technology, the main purpose of LWRs was to generate electricity. Until the minimum NOAK cost scenario, LWRs remained as the cheapest reactor technology. The production of district heat seems to have been of secondary priority. As can be seen in Figure \ref{fig:timeSeries_district_DE} in Appendix \ref{secARes}, competitive technologies for district heating compensate for the reduction of LWR capacities in the lowest cost scenario (No. 1).

\begin{figure}[h!]
\centering
\includegraphics[width=\linewidth]{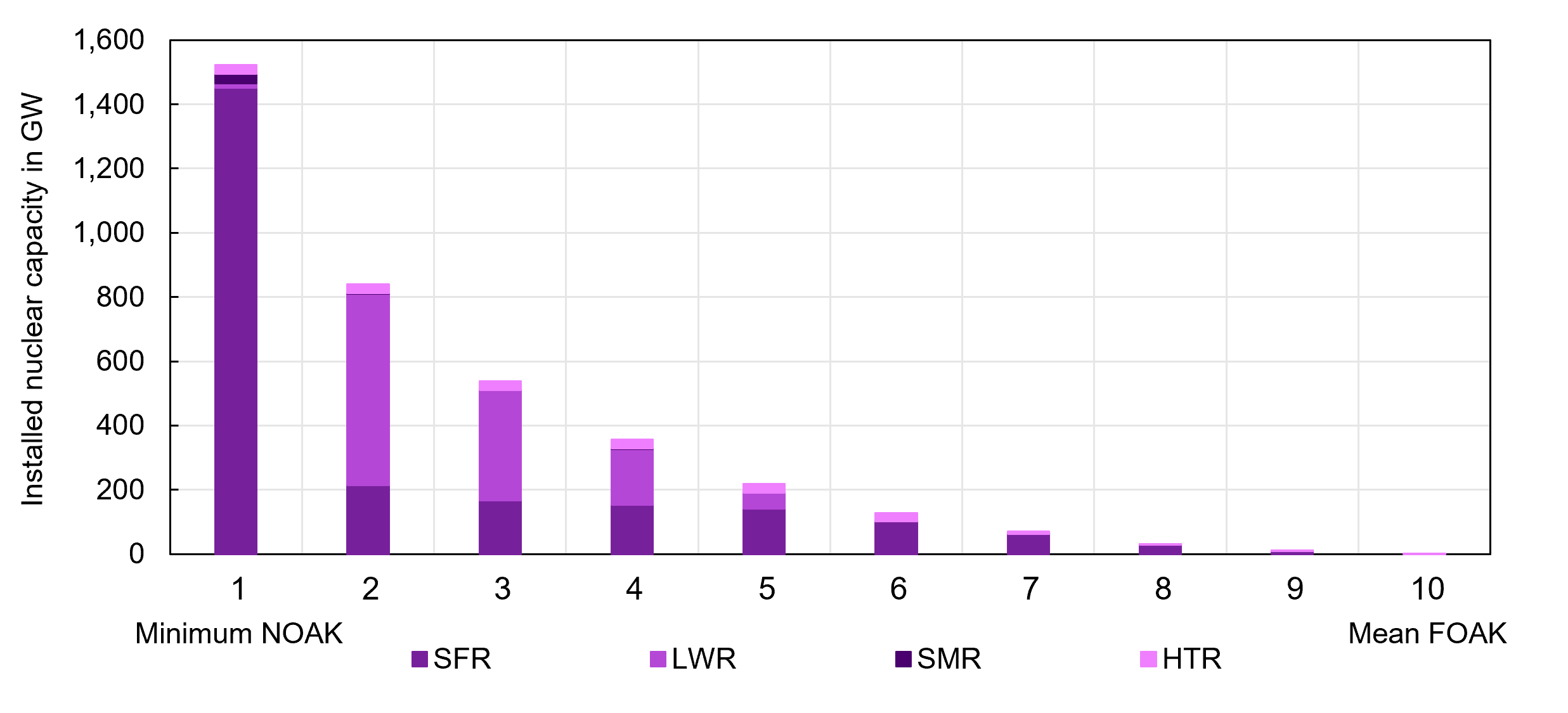}
\caption{Installed nuclear capacities for the different cost scenarios.}
\label{fig:capacities}
\end{figure}

Low-temperature process heat provision remains marginal throughout the scenarios until the lowest cost scenario when SMRs become cheaper than competitive technologies (No. 1). Only then are measurable capacities built. Figure \ref{fig:capacities} shows the nuclear capacities for the different scenarios. The dominance of LWRs for electricity generation is prominent until the lowest cost scenario, when SFRs become cheaper than LWRs (No. 1). Total installed nuclear capacities reach currently installed European capacities (EU and UK, i.e., 102.2 GW \citep{schneider_world_2024}) between scenarios No.~6 \&~7. LWRs are built once their OCC fall enough to compete with alternative electricity generation technologies, i.e., around 3,000 USD/kW (No. 5). But, as mentioned above, LWRs are no longer built once SFRs become even cheaper. In the lowest cost scenario, total nuclear capacities supersede 1,523 GW; driven mainly by very low cost assumptions for SFRs that produce mainly electricity.

As shown in Figure \ref{fig:globalResults}, medium and high process heat provision is swiftly saturated which corresponds to the maximum capacity of HTRs of around 27~GW being reached fairly early (No. 5), and SFR capacity increasing only once the technology becomes cheaper than LWRs (No. 1) (see also Figure \ref{fig:capacities}).

\subsection{System Impact of Nuclear}

Total system costs for the scenarios with low nuclear shares are the highest. However, until installed nuclear capacity reaches levels above 200~GW, these costs remain at an average of €303.6~billion with a standard deviation (sd) of €1.9 billion. In the lowest cost scenario, overall system costs lie at €264.8~billion. Nuclear capacity expansions occur at the expense of open-cycle hydrogen power plants, other gas power plants, solar and wind capacities as well as utility-operated heat plants. Cross-border exchange of electricity is reduced the more nuclear is built. Fuel import dependencies are also reduced, although they remain at similar levels for the last five scenarios.

\begin{figure}[h!]
\centering
\begin{subfigure}{0.9\textwidth}
\includegraphics[width=0.9\linewidth]{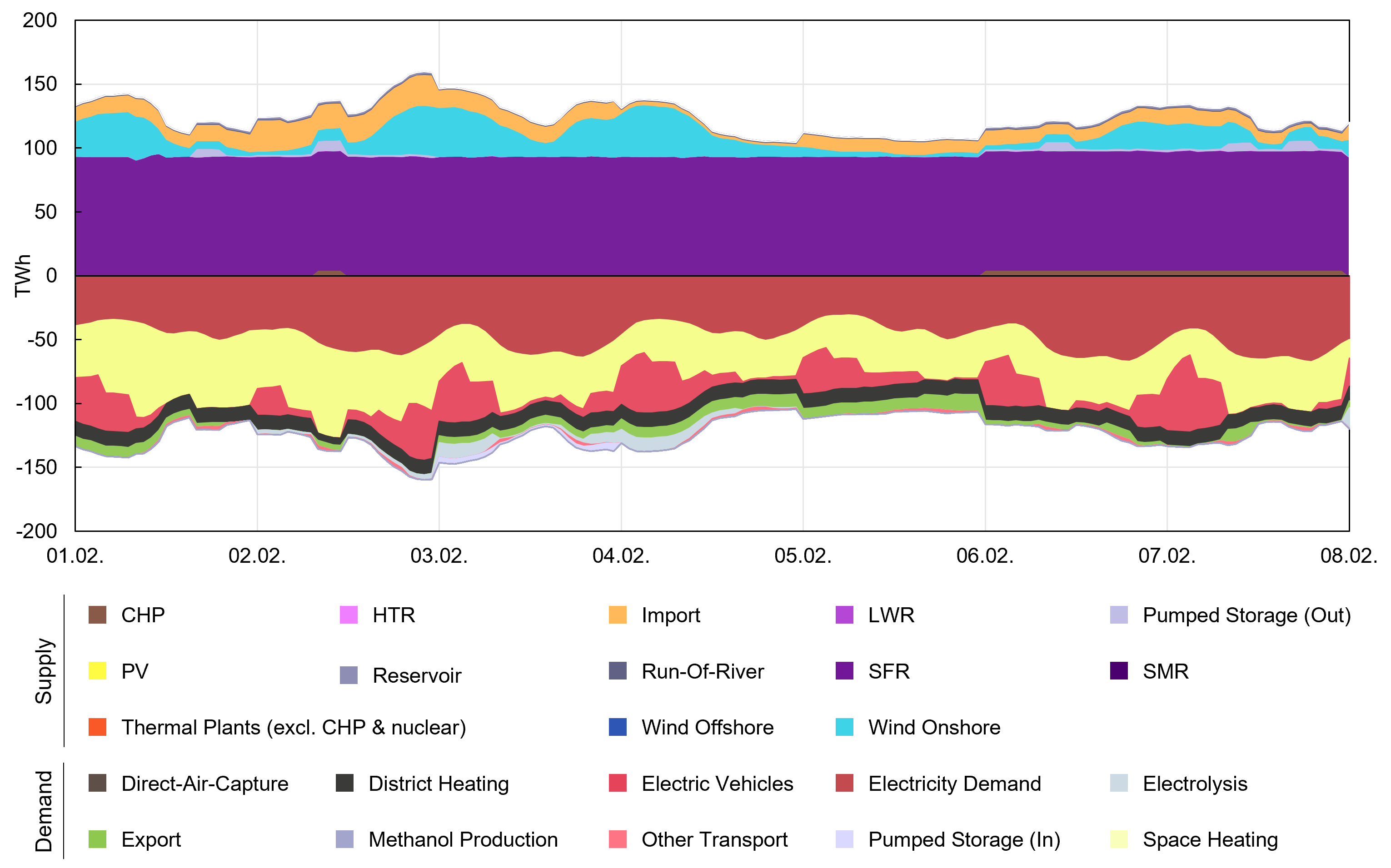}
\caption{Least cost scenario (minimum NOAK, No. 1)}
\label{fig:timeSeries_el_NOAK}
\end{subfigure}
\caption{Exemplary electricity generation in Germany}
\label{fig:timeSeries_el_DE}
\begin{subfigure}{0.9\textwidth}
\includegraphics[width=0.9\linewidth]{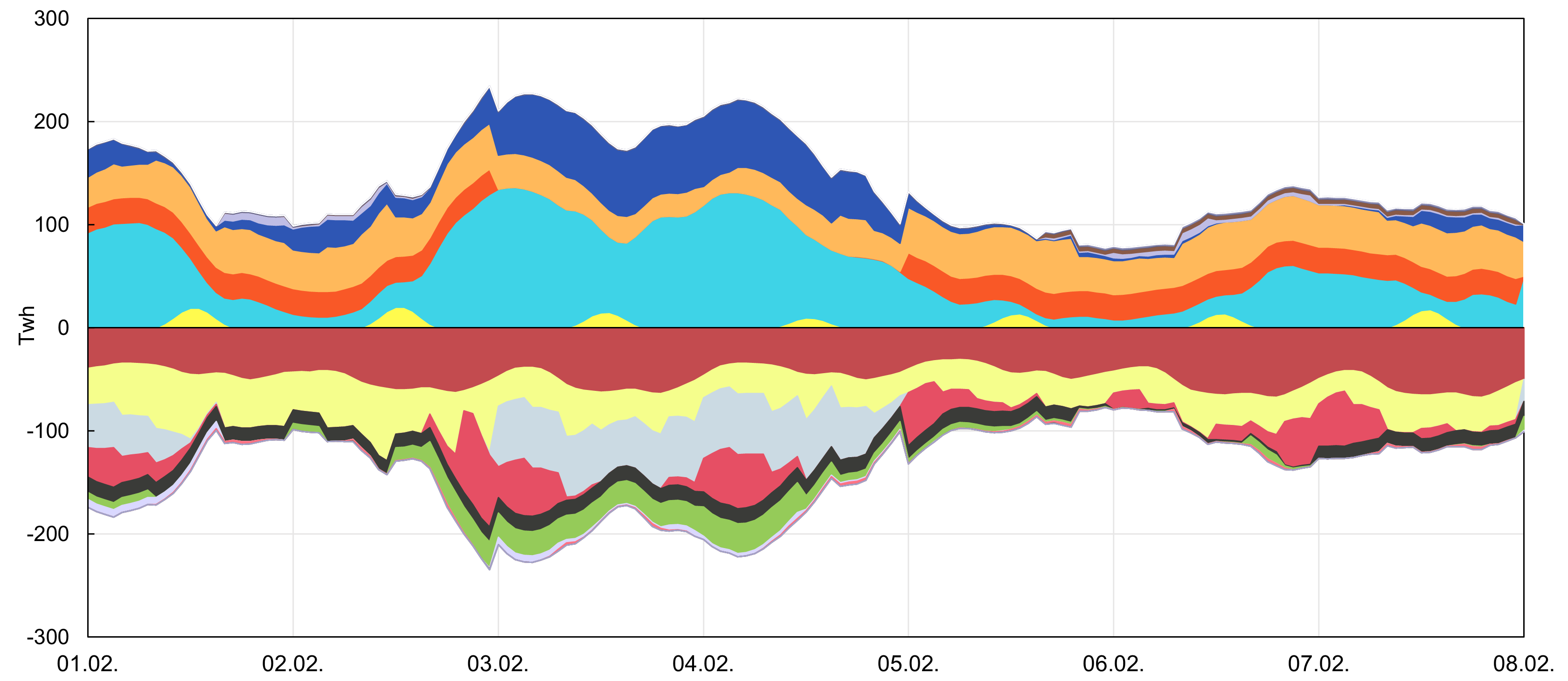} 
\caption{Highest cost scenario (mean FOAK, No. 10)}
\label{fig:timeSeries_el_FOAK}
\end{subfigure}
\end{figure}

Lower system costs are to some extent reflected by the reduced necessity of providing system flexibility for intermittent renewables, such as solar and wind. We assume favorable operational parameters for nuclear, such as a capacity factor of 90\%, that allows the system to run nuclear capacities at almost full capacity. This is shown in Figure \ref{fig:timeSeries_el_NOAK} that depicts a week in February for the German grid in the lowest cost scenario (No. 1). Compared to Figure \ref{fig:timeSeries_el_FOAK}, showing the results for the highest cost scenario (No. 10) and no nuclear in the system, one can see how nuclear blocks the grid instead of complementing intermittent technologies.

\begin{figure}[h!]
\centering
    \begin{subfigure}{0.9\textwidth}
    \includegraphics[width=0.9\linewidth]{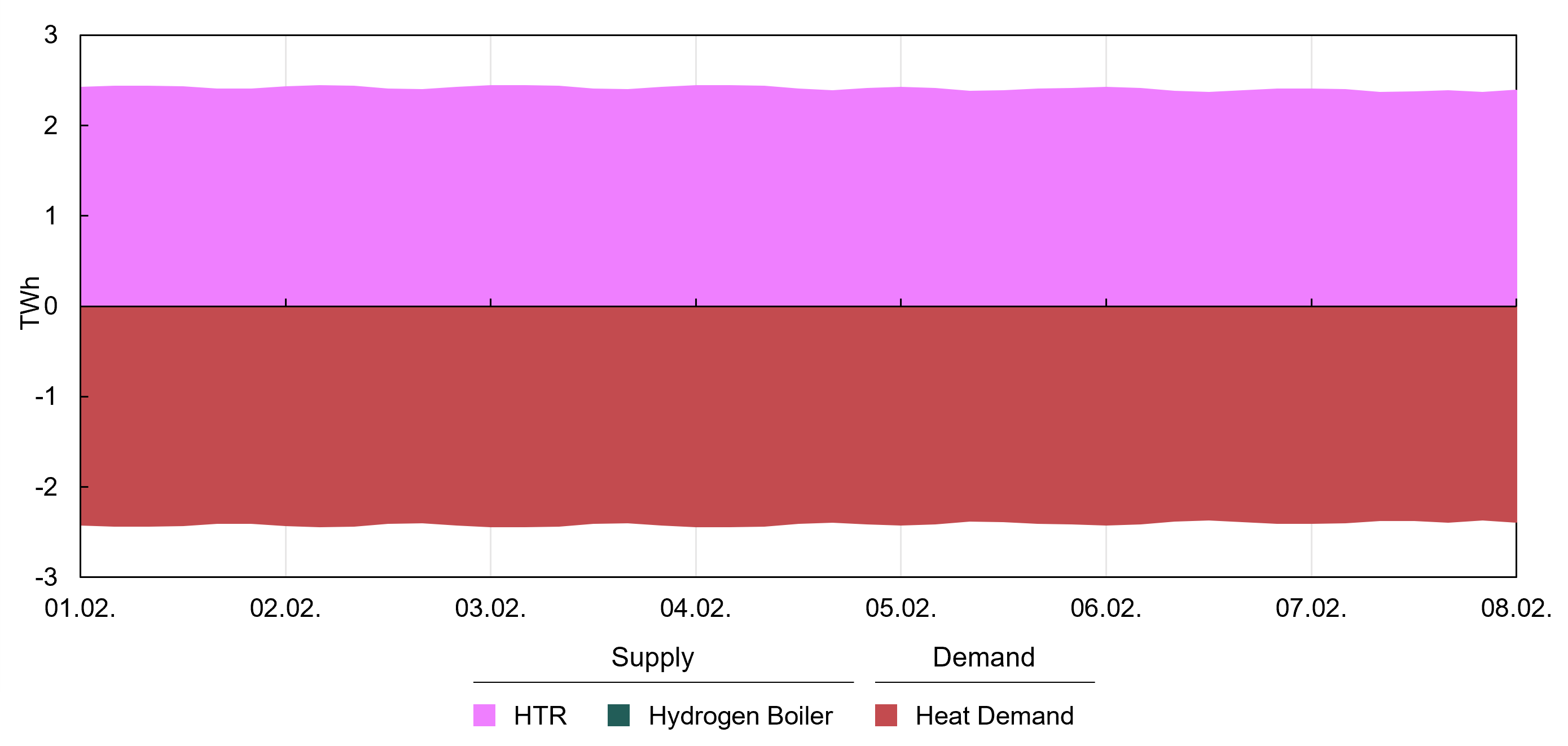}
    \caption{Least cost scenario (minimum NOAK, No. 1)}
    \label{fig:processHeat_high_NOAK}
    \end{subfigure}
    \caption{Exemplary high-temperature process heat generation in Germany}
    \label{fig:timeSeries_processHeat_high_DE}
    \begin{subfigure}{0.9\textwidth}
    \includegraphics[width=0.9\linewidth]{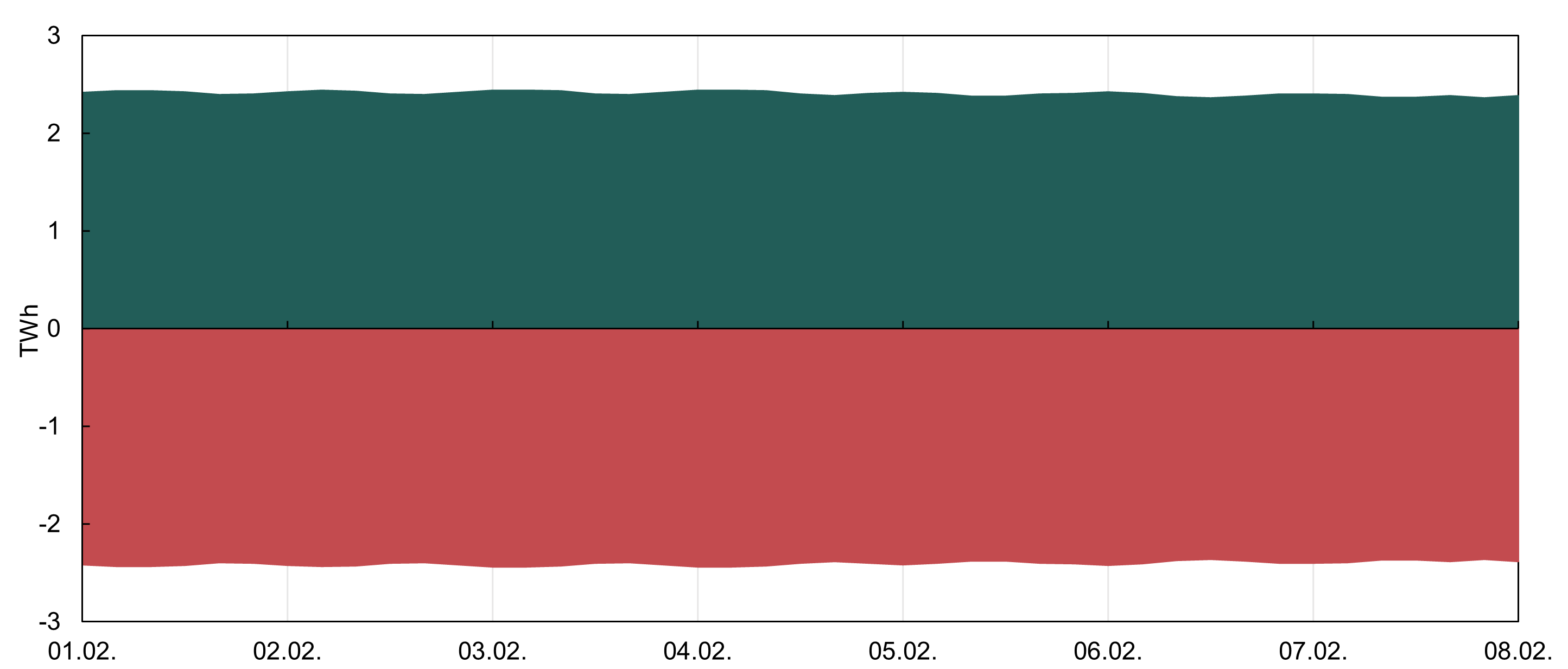} 
    \caption{Highest cost scenario (mean FOAK, No. 10)}
    \label{fig:processHeat_high_FOAK}
    \end{subfigure}
    \end{figure}

A similar pattern occurs for the different types of heat provision.  Heat and electricity, however, are not directly comparable due to inertia, reflected in the model via four-hour instead of hourly timesteps. Regardless, the dominance of nuclear, once built, in heat provision can be observed in Figures \ref{fig:processHeat_high_NOAK}, \ref{fig:processHeat_medium_NOAK}, \ref{fig:processHeat_low_NOAK}, and \ref{fig:districtHeat_FOAK-8}. We show the highest and lowest cost scenarios for each type of heat provision but for simplicity, we only show high process heat generation in this section. Refer to Appendix \ref{secARes} for figures on low and medium process heat as well as district heat. We find that nuclear does not take advantage of the granted flexibility and blocks the grid. This is in line with findings of previous work \citep{goke_flexible_2025}.

\section{Discussion}\label{discussion}

The results underline the lack of cost competitiveness of nuclear power compared to other energy generation technologies. They show that the cost-efficient capacity expansion of especially non-light water reactors would require substantial cost reductions that are, as described in Section \ref{method_literature}, not to be expected in the coming years.

This work assumes that several different reactor concepts will have been fully developed, licensed, tested, commercialized, and produced in vast quantities in the next few years, which, given the current state of project developments, is an optimistic outlook.

However, if the above-mentioned optimistic cost reductions were to be achieved, the results show that non-light water reactor technologies could contribute to energy system decarbonization in Europe in this particular setting (e.g., assuming favorable capacity factors and industry capability, etc.). SFRs and HTRs quickly saturate their heat markets and LWRs are used mainly for electricity production as long as their are cheaper than other (nuclear) technologies. District heat is perceived by the model as useful by product. SMRs compete with other low-carbon technologies for low-temperature heat and are strictly dominated by LWRs for electricity production; they are thus not built in measurable quantities.

But regardless of costs, nuclear reactors feature additional challenges that would need to be overcome for a substantial capacity ramp-up. For example, the heterogeneity of fuel requirements for different SNR concepts corresponds with that of the reactor designs themselves. Many concepts additionally increase proliferation risks because of their requirement of high assay low enriched uranium (HALEU) fuel for which specialized fuel production facilities would have to be funded and constructed, which in turn requires substantial investment and long-term planning and the limitation of industry efforts to a handful of promising reactor designs \cite{committee_on_merits_and_viability_of_different_nuclear_fuel_cycles_and_technology_options_and_the_waste_aspects_of_advanced_nuclear_reactors_merits_2023, krall_nuclear_2022, kim_challenges_2026}. It is uncertain whether this type of infrastructure could be up and running in time for these reactor concepts to play a significant role in decarbonizing the European energy system.

Furthermore, the cost data mentioned above do not include back-end costs of nuclear, such as decommissioning and waste management, or social costs, such as risks relating to accidents, proliferation, or uranium mining and tilling \cite{ipcc_climate_2014, wheatley_reassessing_2016, sovacool_risk_2014, wimmers_reintegrating_2024}. Additional waste challenges arise from implementing non-light water reactors that generate different kinds of waste from, e.g., HALEU fuel, creating additional and currently unstudied waste streams that would have to be dealt with \cite{krall_nuclear_2022}. However, with most nuclear waste responsibility lying with the respective governments \cite{bell_fixing_2022, barenbold_decommissioning_2024}, substantial funding uncertainties could be removed from projects \cite{bowring_federal_1980}, as suggested to be implemented by \citet{sawicki_role_2021}. This would, however, require substantial political power to be achieved, especially in Europe. Additionally, long-term financial assurance and commitment to a limited number of reactor designs are necessary \cite{brown_engineering_2022}.


Additionally, the assumed flexibility of nuclear power plant operations without constraints can be contested. \citet{lynch_nuclear_2022} demonstrate that today's LWRs can be an effective technology for load following, and \citet{jenkins_benefits_2018} show in their model for the South-West U.S. that flexible nuclear operations can reduce system costs and increase reactor operator revenues. \citet[p. 281]{loisel_load-following_2018} even claim that "in practice, countries with [...] high intermittent renewables [...] need [nuclear power plants] to operate". However, despite theoretical technical feasibility, current nuclear reactors' ability to run flexibly is seldom applied today, with operational experience limited mostly to France \cite{schneider_world_2023}. In the U.S., operational experience of flexibly dispatchable nuclear power plants has been limited despite regulations allowing for flexible operations and load following \cite{monitoring_analytics_llc_state_2023}. Reactors designed to operate as flexible backup capacities in energy systems based mainly on renewables face challenges of long standstills and fast ramp-ups. Running at high capacity factors generating heat for other purposes and then flexibly switching to power generation when required will be technologically challenging to implement \cite{richards_economic_2021, ramana_small_2021}.

Our modelling approach faces some inevitable limitations that could limit the applicability of the results. First, the potential locality of heat demand is not accounted for. This is relevant because, for direct heat provision, sources must be located close to off-takers to minimize losses during transfer. This could be especially relevant for nuclear reactors as these technologies may face substantial opposition from local communities.
Secondly, we assume that reactors can switch between operating modes without delay and incurred cost. Adding this to the high capacity factors of 90\% is an assumption that is advantageous to nuclear. Whether non-LWR concepts can operate at such high capacity factors and operate flexibly remains highly uncertain given the lack of precedence in the historical track record of such concepts, see Section \ref{availability}.

Regarding cost assumptions, we show that OCC must be drastically reduced for nuclear power to be competitive with renewable technologies. In our approach, we apply a uniform cost assumption for nuclear power that does not consider cost differences between vendors, financing schemes, or local cost differences that could occur between, for example, projects in Finland and Hungary or the U.K.

The applied model to compute a cost-efficient share of the different nuclear power plants corresponds to or exceeds the state-of-the-art in energy planning. Inevitable limitations of the model impose a positive bias on nuclear power, so our results constitute an upper bound on the cost-efficient share. Most importantly, we neglect the operational constraints of nuclear power and do not impose restrictions on downtimes or production gradients to keep the underlying optimization problem computationally tractable. Of all considered technologies, these constraints are most relevant for nuclear, but according to previous literature, they are acceptable to ignore, especially if a model includes other short-term flexibility, like batteries or demand-side management \citep{poncelet_impact_2016,poncelet_unit_2020}. 

\section{Conclusion}\label{conclusion}

Our literature analysis shows the various potential non-electrical use cases for nuclear reactors and associatated costs for discussed reactor designs. We show that, as of today, these applications are limited to individual reactors supplying heat to small localized demand centers. Most heat provision experience has been gained in Eastern Europe, where a handful of reactors have provided district heat to neighboring towns, and to Japan, where some reactors service desalination plants. Consequentially, the industry's experience in non-electrical use cases is limited, and propositions of localized heat provision should consider the challenges regarding the implementation of sufficient emergency planning zones, flexible operations, and the availability of potentially cheaper alternatives.

Regarding potential technologies, we find that as of today, both SMR and non-LWR concepts cannot fulfill the prerequisites defined by the \citet{committee_on_laying_the_foundation_for_new_and_advanced_nuclear_reactors_in_the_united_states_laying_2023}. We demonstrate that none of the discussed concepts is commercially available. Most concepts are in pre-licensing or licensing stages. However, with ongoing research and development efforts, first concepts might become available towards the turn of the decade. Regardless, the remaining challenges of becoming competitive and affordable (yet socially acceptable) remain to be proven.

Despite the lack of empirical data, many projections and assumptions exist regarding the future cost of diverse reactor concepts. The discrepancy between FOAK and NOAK cost estimations is substantial, and the range of cost estimations in a given readiness level is, too. Notably, NOAK cost estimations are referenced more often, albeit the exact number "n" of required produced units to reach these costs through economies of multiples remains undefined. Compared to ongoing reactor construction projects in OECD countries, the cost assumptions are highly optimistic. Furthermore, as cost overruns and construction delays have historically characterized the nuclear power industry, costs for actual reactor projects are likely to be higher.

Nonetheless, we assess the potential application of LWR and non-LWR concepts in a decarbonized near-future energy system in Europe. In ten scenarios, we continuously reduce OCC values for four different reactor types. We find that nuclear capacities are built primarily for energy types whose provision requires similarly expensive technologies, i.e., medium and high temperature process heat. The required cost ranges seem rather unlikely given the early stages of development for most non-LWR and SMR concepts as well as past experiences from high-capacity LWR construction projects in Europe.


The results described above give an optimistic upper-bound of the potential role of nuclear power in a future energy system because the model's assumptions are skewed positively towards the considered reactor technologies. This includes optimistic assumptions of 90\% capacity factors, the assumption that reactors can be built at NOAK costs within a few years, full operational flexibility of reactors and the ability to switch between electricity and heat provision without ramp-up restrictions, as well as the neglect of decommissioning and waste management costs, potential safety challenges of non-LWR concepts, proliferation risks, and the uncertainty regarding actual industry capability to deliver the required capacities.

Consequentially, policymakers must address these unresolved industrial challenges of non-LWR concepts before a large-scale implementation is possible. Additionally, open questions regarding the organization of new waste disposal routes and potentially increased proliferation and safety risks must be addressed.


\section*{Acknowledgments}
The authors with to thank the participants of the 2025 IEWT Conference in Vienna, of the 2024 International IAEE Conference in Istanbul and of the 2025 AT-OM Day held in Berlin for helpful comments on earlier versions of this work. We extend our gratitude to Christian von Hirschhausen for his suggestions for the improvement of this work.

\section*{Funding}
The research leading to these results has received funding from the Deutsche Forschungsgemeinschaft (DFG) under project number 423336886. We thank the University Library of TU Berlin for providing funds for an open access publication.

\section*{Declaration of Interest}
The authors declare that they have no known competing financial interests or personal relationships that could have appeared to influence the work reported in this paper.

\section*{Data Availability}
All data used in this research is listed either in the Appendix or will be made available upon request. Model code and input data can be accessed via \url{https://github.com/leonardgoeke/EuSysMod/tree/greenfield_nuclearHeat}. Due to their size, model outputs are not part of the repository but are available upon request.
The applied model uses the open AnyMOD.jl modeling framework \citep{goke_anymodjl_2021}. The applied version of the tool is openly available under this link: \url{https://github.com/leonardgoeke/AnyMOD.jl/releases/tag/flexibleElectrificationWorkingPaper}.

\section*{Author Contributions}
A.W.: Conceptualization; Methodology; Formal analysis; Investigation; Data Curation; Writing - Original Draft; Writing - Review \& Editing; Visualization; F.B.: Investigation; Validation; Writing - Review \& Editing; L.G.: Methodology; Validation; Software; Writing - Review \& Editing

\printcredits

\bibliographystyle{unsrtnat}
\bibliography{anyatom_references}

\appendix

\begin{landscape}
\section{Heat Provision Examples}\label{secA2}

\footnotesize
\begin{longtable} {p{1cm}p{2cm}p{1cm}p{2cm}p{2cm}p{2cm}p{2cm}p{2cm}}
    \caption{Non-electrical applications of nuclear power plants. Reactor data (capacity, type and operational status) were taken, where not otherwise specified, from \cite{iaea_power_2024}. Note: *Suspended operation; **Suspended operation from 1997 to 2012; ***Suspended operation from 1995 to 2012; ****Suspended operation from 1998 to 2004; *****Suspended operation from 1998 to 2003}
\label{tab:nuclear_applications}\\
    \hline
    \textbf{Country} & \textbf{Reactor}                                    & \textbf{Type} & \textbf{Application}        & {\textbf{Heat provided (MWth)}} & \multicolumn{1}{l}{\textbf{Operational Lifetime (Reactor)*}} & {\textbf{Gross Power (MWe)}} & \textbf{Reference} \\
    \hline
    \endfirsthead
    \caption[]{(continued)}\\
    \hline
    \textbf{Country} & \textbf{Reactor}                                    & \textbf{Type} & \textbf{Application}        & {\textbf{Heat provided (MWth)}} & {\textbf{Operational Lifetime (Reactor)*}} & {\textbf{Gross Power (MWe)}} & \textbf{Reference} \\ \hline
    \endhead
    \hline
    \multicolumn{8}{l}{Continued on next page.}\\
    \hline
    \endfoot
    \hline
    \endlastfoot
Bulgaria         & Kozloduy-1                                          & PWR           & District heat               & 230 (total)                                       & 1974 - 2002                                                  & 440                                            & \cite{kupitz_small_2001}                                                                                                                     \\
         & Kozloduy-2                                          & PWR           & District heat               & 230 (total)                                       & 1975-2002                                                    & 440                                            & \cite{kupitz_small_2001}                                                                                                                      \\
         & Kozloduy-3                                          & PWR           & District heat               & 230 (total)                                       & 1981-2006                                                    & 440                                            & \cite{kupitz_small_2001}                                                                                                                    \\
         & Kozloduy-4                                          & PWR           & District heat               & 230 (total)                                       & 1982-2006                                                    & 440                                            & \cite{kupitz_small_2001}                                                                                                                      \\
\hline
\end{longtable}

\end{landscape}

\section{Non-Light Water Reactor Technologies}\label{secA1}

In this section of the appendix, we provide a brief technological overview of the non-light water reactors discussed in Section \ref{availability}. Other concepts, not discussed here, include Accelerator-Driven Systems \cite{biarrotte_beam_2013}, and Supercritical water-cooled reactors \cite{pistner_analysis_2024}. In this overview, we also give examples of existing or planned experimental, demonstration, and, if available, commercial reactors.

\paragraph{Sodium Cooled Fast Reactors (SFR)}
SFRs operate on a fast neutron spectrum and thus require no moderator. The core is cooled by liquid sodium. SFRs can operate on Mixed-oxide (MOX) fuel, which contains plutonium, or on enriched uranium fuel. They output temperature ranges from 400 to 450 °C. The main technical challenges SFRs have faced in the past relate to the high reactivity of sodium when in contact with water and air. Furthermore, the opaque nature of liquid sodium complicates inspection, monitoring and fuel exchange operations and induces component stress. To establish a fleet of SFRs, fuel manufacturing infrastructure must be set up, and additional proliferation safeguarding measures will have to be set up. Potential applications, in addition to electricity and heat provision, is the potential of so-called waste "recycling" via reprocessing facilities (to be reestablished) and plutonium breeding in the reactor itself \cite{pistner_analysis_2024,schulenberg_fourth_2022, cochran_fast_2010}. 
SFR research has been ongoing since the 1940s, with the first experimental reactors built in the 1950s and 1960s, e.g., EBR-1, Fermi-1, and EBR-2 in the U.S., DFR in the U.K., or Rapsodie in France. All of these reactors are closed. In the 1970s, further experimental and demonstration reactors were built in Germany (KNK-II (closed), SNR-300 (never critical)), France (Phénix and Superphénix (both closed)), the Soviet Union/Kazakhstan (BN-350 (closed)), the U.K. (PFR, closed), Japan (Jōyō (operational), Monju (closed)), and India (FBTR (operational)). China commissioned its first experimental SFR CEFR in 2012. The only two operational commercial SFRs, BN-600 and BN-800, are located in Russia \cite{schneider_world_2024,ohshima_sodium-cooled_2016,pistner_analysis_2024,bose_questioning_2024}.
Two SFRs are currently under construction in India (PFBR) and China (CFR-600). Other ongoing development projects are in various planning and licensing stages. These include the Indian FBR 1 \& 2 projects, the Japanese Toshiba 4S, the Russian BN-1200 and BOR-60 (MBIR), the South Korean PGSFR, and GE Hitachi's PRISM and VTR, as well as TerraPower's TWR, both in the U.S.. The French ASTRID project was discontinued in 2019 \cite{iaea_status_2022,ohshima_sodium-cooled_2016,schulenberg_fourth_2022}.

\paragraph{Lead-cooled fast reactors (LFR)}
LFRs operate with fast neutrons and thus require no moderator. The coolant is liquid lead or lead-bismuth composite. LFRs are planned to run on MOX fuel with output temperatures of 400 to 620°C. The main challenges for LFR implementation relate to the supply of MOX fuel and required infrastructure, the management of generated polonium-210, corrosion and material erosion risks because fuel rods containing iron, nickel and cobalt would be diluted in the coolant, the sensitivity of a potential reactor to seismic activity due to the high density and mass of the coolant, and finally, inspection and monitoring challenges due to opaque coolant. Historically, LFRs were used for marine propulsion, but in addition to the same potential applications of SFRs, they could provide load-following services due to the theoretically short ramp-up times \cite{pistner_analysis_2024, alemberti_lead_2021, schulenberg_fourth_2022, soto_dispatch_2022}.
Initial development efforts began in 1942 in the U.S. but were soon canceled. The first experimental reactor 27/VT was built in 1959 in the Soviet Union for marine propulsion. In the 1970s, seven Soviet Lira-class submarines were fitted with LFRs, all of which have been out of service since the mid-1990s \cite{rawool-sullivan_technical_2002}. Currently, there is no operational LFR. Current concepts are the Belgian MYRRHA (also planned as an accelerator-driven system), the Chinese CLFR-100, CLFR-300 and BLESS, the EU-funded projects ALFRED and ELFR, the LRF-AS-200 under development in Luxemburg, the Russian BREST-OD-300 (under construction) and SVBR-100, the South Korean PEACER, and the Westinghouse-LFR in the U.S. \cite{alemberti_lead_2021,schulenberg_fourth_2022,pistner_analysis_2024}.

\paragraph{Gas-cooled fast reactors (GFR)}
In addition to SFRs and LFRs discussed in the main text, there exist fast reactor concepts not based on liquid metal coolants, but instead on gasses, such as Helium or CO\textsubscript{2}. Depending on the coolant, these GFRs can reach output temperatures ranging from 500 to 850°C. Up to 600°C, MOX fuel could be used for operations, at higher temperatures, to-be-developed uranium-plutonium-carbide fuels could be applied. Technical challenges are the low thermal conductivity of gasses, complicating cooling system design and inflating the size of the reactor itself, the unavailability of required fuels, and the availability of suitable materials capable of whithstanding high temperatures and pressures. The main advantage compared to SFRs and LFRs is the low reactivity of helium (inert), and the potentially high thermal efficiency. Furthermore, the high output temperatures could be used to supply valuable process heat to the industry \cite{hatala_gas_2021, tsvetkov_gas-cooled_2016}.
The first reactor concept was designed around 1962 by General Atomics in the U.S., followed by EU-funded concept designs from the 1970s onwards (GBR 1-4), and the Enhanced Gas-Cooled Reactor concept in the U.K., designed in the 1990s. However, no reactor was built. Since the early 2000s, France has been fostering GFR research, leading to the proposed ALLEGRO concept. Other concepts under development are the  KAMADO FBR in Japan and the General Atomics EM$^{2}$ \cite{gif_gas-cooled_2024,hatala_gas_2021,tsvetkov_gas-cooled_2016}.

\paragraph{Molten Salt Reactors (MSR)}
The term MSRs comprises a multitude of different reactor concepts and designs. The common denominator is the chosen coolant being a type of molten salt, based on lithium or sodium in combination with fluoride or chloride. Furthermore, some concepts run on fast, some on slow neutron spectra. The latter require graphite as moderator. Fuel can be sometimes integrated into the molten salts, and be of various liquid and solid concepts, such as HALEU, uranium-plutonium, uranium-thorium, or plutonium combined with minor actinides. Output temperatures range around 600 to 700°C. In addition to heat and electricity provision, MSRs could function as thermal storage facilities \citep{soto_dispatch_2022}. Current challenges limiting MSR establishment relate to the identification and testing of suitable salt-material combinations and the development of integrated models for safety case assessments. Furthermore, the radioactive salts require the establishment of infrastructure for storage and transport. Additionally, proliferation and safety regulations might have to be adapted \cite{pistner_analysis_2024, riley_molten_2019}.
The U.S. Air Force first researched MSRs for aircraft propulsion in the 1950s, followed by the MSRE experimental reactor built at the Oak Ridge National Laboratory (ORNL) in the 1960s. This reactor has been in a long-term enclosure since 1969 \cite{aec_evaluation_1972}. There are no currently operating MSRs, but an array of different concepts are under development. Thermal fluoride-based MSRs with solid fuel are under development in China (TSMR-SF) and the U.S. Here, the ORNL is working on the AHTR and Mark 1 PB-FHR concepts, and Kairos Power is developing the KP-FHR. Thermal fluoride-based MSRs with liquid fuel are under development in Canada (Terrestrial Energy's IMSR-400), China (TSMR-LF), Denmark (Seaborgy Technology's MSTR or CMSR concept), Japan (FUJI), and the ORNL's ThorCon in the US. Two projects, the Russian MOSART and the internationally-funded MSFR (funded by Australia, Canada, the EU, France, Russia, Switzerland, and the U.S.) are concepts for fast fluoride-based MSRs with liquid fuel. Finally, three fast chloride-based MSR concepts with liquid fuel are under development. The projects are the German-designed, recently Canada-based Dual-Fluid reactor, and US-based concepts MSCFR by Elysium, and the SSR-W by Moltex Energy \cite{riley_molten_2019,pistner_analysis_2024}.

\paragraph{(Very-)High-temperature reactors (HTR)}
HTRs are moderated with graphite and require gas, such as helium, as coolant. They require so-called TRISO fuel, usually pebble-based, to operate. This fuel can be based on uranium, plutonium, or thorium. The advantage of HTRs is their potentially very high output temperature of up to 1,000°C (albeit experimental reactors have yielded only 800°C), making them promising for providing high-temperature process heat for, e.g., coal gasification or cement production. Depending on the design, they might be suitable for reprocessed fuel, and high operating temperatures increase thermal efficiency. However, these high temperatures pose the most significant challenges for HTR development. For example, graphite moderation functions only up to 950°C. Thus new materials are required for higher temperatures. Every component must be able to withstand high corrosion and stress. Further, operations must ensure that internal temperatures do not reach the TRISO fuel melting point at 1,800°C. Additionally, waste volumes are inflated due to pebble-based fuel and graphite \citep{ramana_checkered_2016,pistner_analysis_2024,schultz_large-scale_2003, moormann_caution_2018}.
The first experimental reactors were built in the 1960s in the UK (Dragon) and the US (Peach Bottom and AVR); they are all closed. Two demonstration reactors built in the 1980s in Germany (THTR) and the U.S. (Fort St. Vrain) have also been closed. Two experimental reactors are operational in China (HTR-10), and Japan (HTTR). China also operates the only commercial-scale HTR worldwide, the 210-MW-HTR-PM, connected to the grid in 2021. Current concepts under development are the Japanese GTHTR300, the Polish TeResa, and several concepts in the U.S., such as the FHR, the AHTR-100, and X-Energy's Xe-100 microreactor. The ANTARES and Pismatic Modular HTGRs were both discontinued \citep{zhang_shandong_2016, buongiorno_future_2018, bose_questioning_2024,ramana_checkered_2016, abou-jaoude_literature_2023}.

\section{Cost and Technology Data}\label{secA3}

This section of the appendix provides all the data from the literature analysis. Tables \ref{tab:occ_data}, \ref{tab:om_fix_data}, and \ref{tab:om_var_data} give the minimum, mean, and maximum OCC and OM cost normalized to USD-2019, as well as the number of retrieved data points, per reactor type. Table \ref{tab:inflation} shows the factors used to normalize the cost data taken from the literature shown in Tables \ref{tab:cost_raw_occ}, \ref{tab:cost_raw_om_fix} and \ref{tab:cost_raw_om_var}.

\begin{table}[h]
\begin{tabular}{llrrrr}
\hline
\textbf{Readiness Level} & \textbf{Reactor Type} & \multicolumn{1}{l}{\textbf{Data Points}} & \multicolumn{1}{l}{\textbf{Mean}} & \multicolumn{1}{l}{\textbf{Min}} & \multicolumn{1}{l}{\textbf{Max}} \\ \hline
FOAK                     & LWR                   & 6                                               & 5,622.9                                            & 3,880                                              & 7,164                                              \\
                         & SMR                   & 16                                               & 9,310.45                                            & 3,512.84                                           & 24,309.00                                             \\
                         & SFR                   & 5                                               & 9,511.47                                           & 4,641.00                                           & 20,511.00                                             \\
                         & HTR                   & 18                                              & 9,438.52                                            & 3,466.2                                            & 26,452.22                                          \\
                         & LFR                   & 3                                              & 6,545.00                                            & 5,355.00                                            & 8,449.00                                          \\
NOAK                     & LWR                   & 20                                              & 4,979.99                                            & 1,782.62                                            & 9,035.40                                            \\
\hline
\end{tabular}
\caption{OCC data for reactor types and readiness levels in USD-2019 per kW. Abbreviations used: LWR = High-capacity light water reactor; SMR = Small Modular Reactor (limited to light water technology); SFR = sodium-cooled fast reactor; HTR = High-temperature reactor; LFR = lead-cooled fast reactor; MSR = molten-salt reactor (thermal); MSFR = molten salt fast reactor}
\label{tab:occ_data}
\end{table}

\begin{table}
\begin{tabular}{llrrrr}
\hline
\textbf{Readiness Level} & \textbf{Reactor Type} & \multicolumn{1}{l}{\textbf{Data Points}} & \multicolumn{1}{l}{\textbf{Mean}} & \multicolumn{1}{l}{\textbf{Min}} & \multicolumn{1}{l}{\textbf{Max}} \\ \hline
    FOAK                     & LWR                   & 3                                               & 175.96                                            & 38.80                                              & 398.00                                              \\
                         & SMR                   & 1                                               & 175.96                                            & 175.96                                           & 175.96                                             \\
                         & SFR                   & 1                                               & 398                                           & 398                                           & 398                                             \\
                         & HTR                   & 9                                              & 198.42                                            & 20.18                                            & 613.04                                          \\
NOAK                     & LWR                   & 3                                              & 124.07                                            & 31.00                                            & 199.00                                            \\
                         & SMR                   & 1                                               & 95.00                                            & 95.00                                              & 95.00                                              \\
\hline
\end{tabular}
\caption{Fixed operation and maintenance cost data for reactor types and readiness levels in USD-2019 per kW. Abbreviations used: LWR = High-capacity light water reactor; SMR = Small Modular Reactor (limited to light water technology); SFR = sodium-cooled fast reactor; HTR = High-temperature reactor; LFR = lead-cooled fast reactor; MSR = molten-salt reactor (thermal); MSFR = molten salt fast reactor}
\label{tab:om_fix_data}
\end{table}

\begin{table}
\begin{tabular}{llrrrr}
\hline
\textbf{Readiness Level} & \textbf{Reactor Type} & \multicolumn{1}{l}{\textbf{Data Points}} & \multicolumn{1}{l}{\textbf{Mean}} & \multicolumn{1}{l}{\textbf{Min}} & \multicolumn{1}{l}{\textbf{Max}} \\ \hline
    FOAK                     & LWR                   & 2                                               & 3.6                                            & 2.36                                              & 4.85                                              \\
                         & SMR                   & 2                                               & 25.17                                            & 21.34                                           & 29                                             \\                        
                         & HTR                   & 4                                              & 20.12                                            & 0.34                                            & 30.81                                          \\
NOAK                     & LWR                   & 2                                              & 4.71                                            & 4.44                                            & 4.98                                            \\
                         & SMR                   & 2                                               & 8.78                                            & 3                                              & 2.33                                              \\
                         & SFR                   & 1                                              & 2.33                                            & 2.33                                              & 2.33                                              \\
\hline
\end{tabular}
\caption{Variable operation cost data for reactor types and readiness levels in USD-2019 per MWh. Abbreviations used: LWR = High-capacity light water reactor; SMR = Small Modular Reactor (limited to light water technology); SFR = sodium-cooled fast reactor; HTR = High-temperature reactor; LFR = lead-cooled fast reactor; MSR = molten-salt reactor (thermal); MSFR = molten salt fast reactor}
\label{tab:om_var_data}
\end{table}

\begin{table}
\begin{tabular}{lrr} \hline
Year & Inflation to 2019 & Combined Cost Factor following Abou-Jaoude et al. (2023) \\ \hline
1978 & 392\%             & 6.6                                                      \\
1979 & 352\%             & 6.22                                                     \\
1980 & 310\%             & 5.9                                                      \\
1981 & 281\%             & 5.35                                                     \\
1982 & 265\%             & 4.89                                                     \\
1983 & 257\%             & 4.5                                                      \\
1984 & 246\%             & 4.17                                                     \\
\hline 
\end{tabular}
\caption{Inflation rates to 2019 taken from inflationtool.com and cost escalation factors chosen following \citet{abou-jaoude_literature_2023}.}
\label{tab:inflation}
\end{table}

\begin{landscape} 
\footnotesize
\begin{longtable} {p{2cm}p{1cm}p{2cm}p{2cm}p{1cm}p{1cm}p{3cm}p{1cm}p{2cm}}
    \caption{OCC Data from literature analysis for different reactor types and (assumed) readiness levels. Abbreviations: SFR = Sodium-cooled fast reactor; LWR-SMR = Light water small modular reactor; HTR = High-temperature reactor; LWR = High-capacity ligh water reactor; HTGR = High temperature gas-cooled reactor; MSR = Molten salt reactor; SMR = Small modular reactor; NGNP = Next generation nuclear power plant; PWR = Pressurized water reactor;}
\label{tab:cost_raw_occ}\\
    \hline
    {\textbf{Reactor}} & {\textbf{Reactor Type}}                                    & {\textbf{Capacity in MWe}} & {\textbf{OCC}}        & {\textbf{Unit}} & {\textbf{Year}} & {\textbf{Description}} & {\textbf{Readiness Level}} & {\textbf{Reference}} \\
    \hline
    \endfirsthead
    \caption[]{(continued)}\\
    \hline
    {\textbf{Reactor}} & {\textbf{Reactor Type}}                                    & {\textbf{Capacity in MWe}} & {\textbf{OCC}}        & {\textbf{Unit}} & {\textbf{Year}} & {\textbf{Description}} & {\textbf{Readiness Level}} & {\textbf{Reference}} \\ \hline
    \endhead
    \hline
    \multicolumn{9}{l}{Continued on next page.}\\
    \hline
    \endfoot
    \hline
    \endlastfoot
 
ASTRID                                          & SFR                                          & 600                                                                & 7167                                & USD/kWe                              & 2010                                 & Total Project   Cost                                                       & FOAK                                            & \cite{rodriguez_astrid_2018, pistner_analysis_2024}                                                                              \\
NuScale                                         & LWR-SMR                                      & 924                                                                & 3600                                & USD/kWe                              & 2018                                 & NOAK OCC                                                                   & NOAK                                            & \cite{rothwell_projected_2022}                                                                          \\
NuScale                                         & LWR-SMR                                      & 924                                                                & 8000                                & USD/kWe                              & 2018                                 & FOAK OCC                                                                   & FOAK                                            & \cite{rothwell_projected_2022}                                                                         \\
\hline
\end{longtable}

\footnotesize
\begin{longtable} {p{2cm}p{1cm}p{2cm}p{2cm}p{1cm}p{1cm}p{3cm}p{1cm}p{2cm}}
    \caption{OM data from literature analysis for different reactor types and (assumed) readiness levels. Abbreviations: SFR = Sodium-cooled fast reactor; LWR-SMR = Light water small modular reactor; HTR = High-temperature reactor; LWR = High-capacity ligh water reactor; HTGR = High temperature gas-cooled reactor; MSR = Molten salt reactor; SMR = Small modular reactor; NGNP = Next generation nuclear power plant; PWR = Pressurized water reactor}
\label{tab:cost_raw_om_fix}\\
    \hline
    \textbf{Reactor} & \textbf{Reactor Type} & \textbf{Capacity in MWe} & \textbf{OM Cost (fix)}        & {\textbf{Unit}} & {\textbf{Year}} & {\textbf{Description}} & {\textbf{Readiness Level}} & \textbf{Reference} \\
    \hline
    \endfirsthead
    \caption[]{(continued)}\\
    \hline
   \textbf{Reactor} & \textbf{Reactor Type} & \textbf{Capacity in MWe} & \textbf{OM Cost (fix)}        & {\textbf{Unit}} & {\textbf{Year}} & {\textbf{Description}} & {\textbf{Readiness Level}} & \textbf{Reference} \\ \hline
    \endhead
    \hline
    \multicolumn{9}{l}{Continued on next page.}\\
    \hline
    \endfoot
    \hline
    \endlastfoot
HTGR                                            & HTR                   & 600                                         & 99.63333333    & USD/kWe       & 2012          & Annual O\&M   Costs                                                        & FOAK                     & \cite{budi_fuel_2019}\\
HTGR                                            & HTR                   & 350                                         & 161.8          & USD/kWe       & 2012          & Annual O\&M Costs                                                          & FOAK                     & \cite{budi_fuel_2019} \\
HTGR                                            & HTR                   & 80                                          & 39             & USD/kWe       & 2020          & Lower Cost,   High Learning (Best Case) // FOAK CAPEX                      & FOAK                     & \cite{stein_advancing_2022} \\
Advanced   reactor with thermal storage (ARTES) & LWR                   & 345                                         & 121            & USD/kWe       & 2020          & Lower Cost, High Learning (Best Case) // FOAK CAPEX                        & FOAK                     & \cite{stein_advancing_2022}  \\
HTGR                                            & HTR                   & 80                                          & 189            & USD/kWe       & 2020          & Upper Cost, Low   Learning (Worst Case) // FOAK CAPEX                      & FOAK                     & \cite{stein_advancing_2022}   \\
Advanced   reactor with thermal storage (ARTES) & LWR                   & 345                                         & 40             & USD/kWe       & 2020          & Upper Cost, Low Learning (Worst Case) // FOAK CAPEX                        & FOAK                     & \cite{stein_advancing_2022}\\
\hline
\end{longtable}

\footnotesize
\begin{longtable} {p{2cm}p{1cm}p{2cm}p{2cm}p{1cm}p{1cm}p{3cm}p{1cm}p{2cm}}
    \caption{OCC Data from literature analysis for different reactor types and (assumed) readiness levels. Abbreviations: SFR = Sodium-cooled fast reactor; LWR-SMR = Light water small modular reactor; HTR = High-temperature reactor; LWR = High-capacity ligh water reactor; HTGR = High temperature gas-cooled reactor; MSR = Molten salt reactor; SMR = Small modular reactor; NGNP = Next generation nuclear power plant; PWR = Pressurized water reactor}
\label{tab:cost_raw_om_var}\\
    \hline
    {\textbf{Reactor}} & {\textbf{Reactor Type}} & {\textbf{Capacity in MWe}} & {\textbf{OM Cost (var)}}        & {\textbf{Unit}} & {\textbf{Year}} & {\textbf{Description}} & {\textbf{Readiness Level}} & {\textbf{Reference}} \\
    \hline
    \endfirsthead
    \caption[]{(continued)}\\
    \hline
    {\textbf{Reactor}} & {\textbf{Reactor Type}} & {\textbf{Capacity in MWe}} & {\textbf{OM Cost (var)}}        & {\textbf{Unit}} & {\textbf{Year}} & {\textbf{Description}} & {\textbf{Readiness Level}} & {\textbf{Reference}} \\ \hline
    \endhead
    \hline
    \multicolumn{9}{l}{Continued on next page.}\\
    \hline
    \endfoot
    \hline
    \endlastfoot
HTGR                                            & HTR                   & 80                                          & 0.35           & USD/MWh       & 2020          & Lower Cost,   High Learning (Best Case) // FOAK CAPEX                      & FOAK                     & \cite{stein_advancing_2022}  \\
Advanced   reactor with thermal storage (ARTES) & LWR                   & 345                                         & 2.43           & USD/MWh       & 2020          & Lower Cost, High Learning (Best Case) // FOAK CAPEX                        & FOAK                     & \cite{stein_advancing_2022} \\
HTGR                                            & HTR                   & 80                                          & 0              & USD/MWh       & 2020          & Upper Cost, Low   Learning (Worst Case) // FOAK CAPEX                      & FOAK                     & \cite{stein_advancing_2022} \\
Advanced   reactor with thermal storage (ARTES) & LWR                   & 345                                         & 5              & USD/MWh       & 2020          & Upper Cost, Low Learning (Worst Case) // FOAK CAPEX                        & FOAK                     & \cite{stein_advancing_2022} \\
Fast Reactors                                   & SFR                   & n.a.                                        & 2.1            & USD/MWh       & 2017          & OCC                                                                        & NOAK                     & \cite{dixon_advanced_2017} \\
Gas-cooled   reactors                           & HTR                   & n.a.                                        & 0              & USD/MWh       & 2017          & OCC                                                                        & NOAK                     & \cite{dixon_advanced_2017} \\
\hline
\end{longtable}

\end{landscape}

\section{Visualization of Model Results for Heat Types}\label{secARes}

We show lowest and highest nuclear cost scenario results from Germany for each heat type. For district heat, we also show the second-to-last cost scenario. For low temperature process heat, we show data from the French grid because the German case does not include any nuclear for this temperature level.

\begin{figure}[h!]
\centering
    \begin{subfigure}{0.9\textwidth}
    \includegraphics[width=0.9\linewidth]{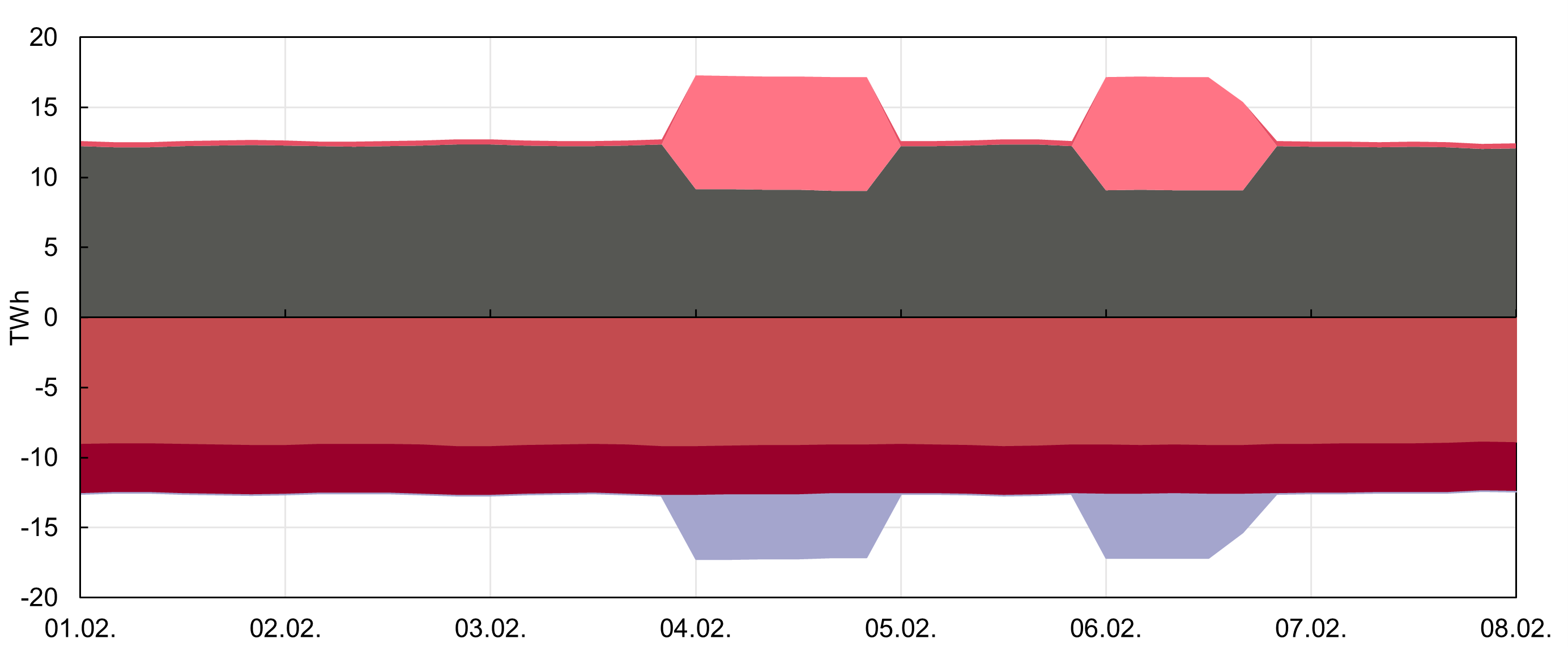} 
    \caption{Highest cost scenario (mean FOAK)}
    \label{fig:processHeat_medium_FOAK}
    \end{subfigure}
    \begin{subfigure}{0.9\textwidth}
    \includegraphics[width=0.9\linewidth]{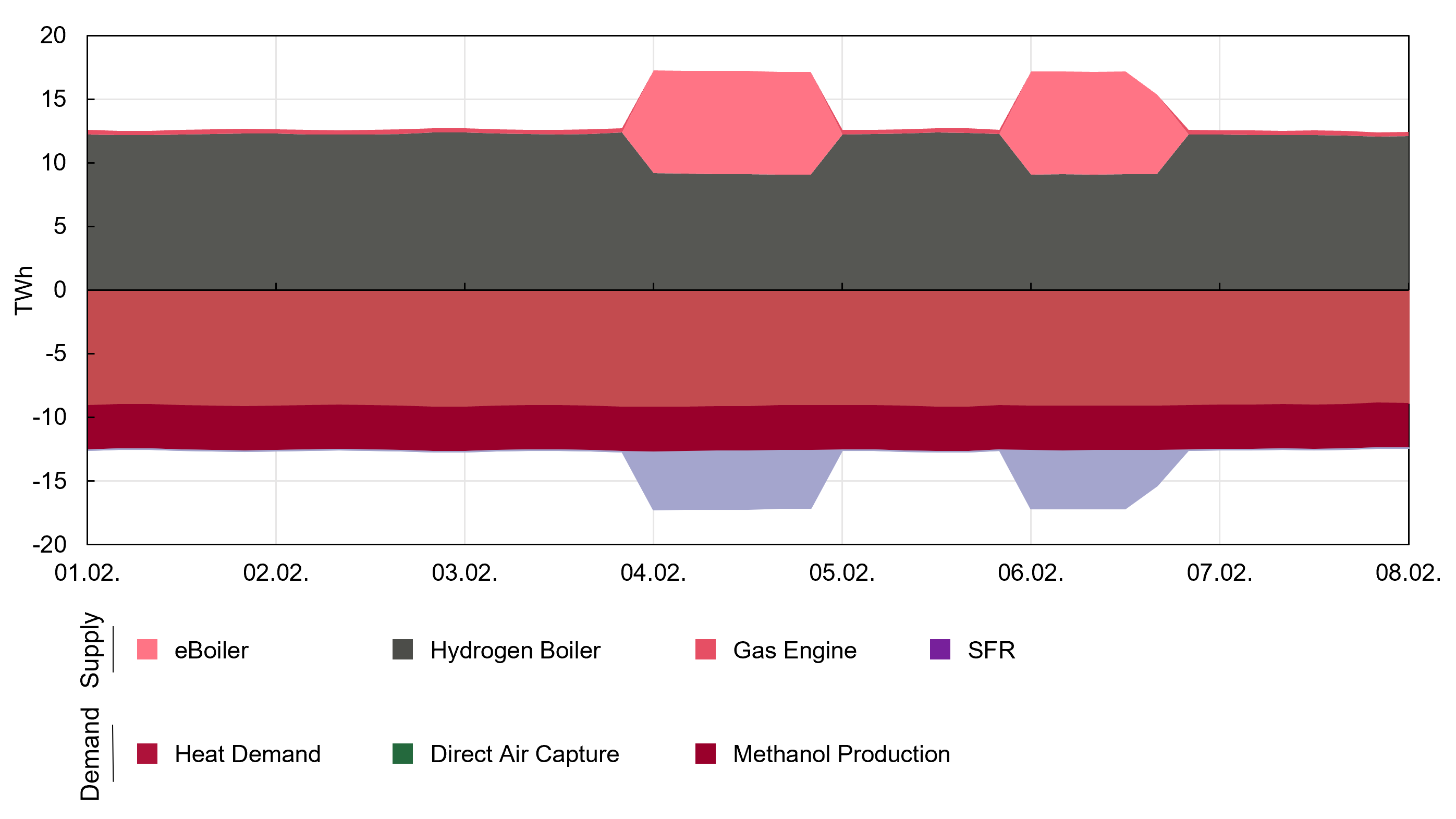}
    \caption{Least cost scenario (minimum NOAK)}
    \label{fig:processHeat_medium_NOAK}
    \end{subfigure}
    \caption{Exemplary medium-temperature process heat generation in Germany}
    \label{fig:timeSeries_processHeat_medium_DE}
\end{figure}

\begin{figure}[h!]
\centering
    \begin{subfigure}{0.9\textwidth}
    \includegraphics[width=0.9\linewidth]{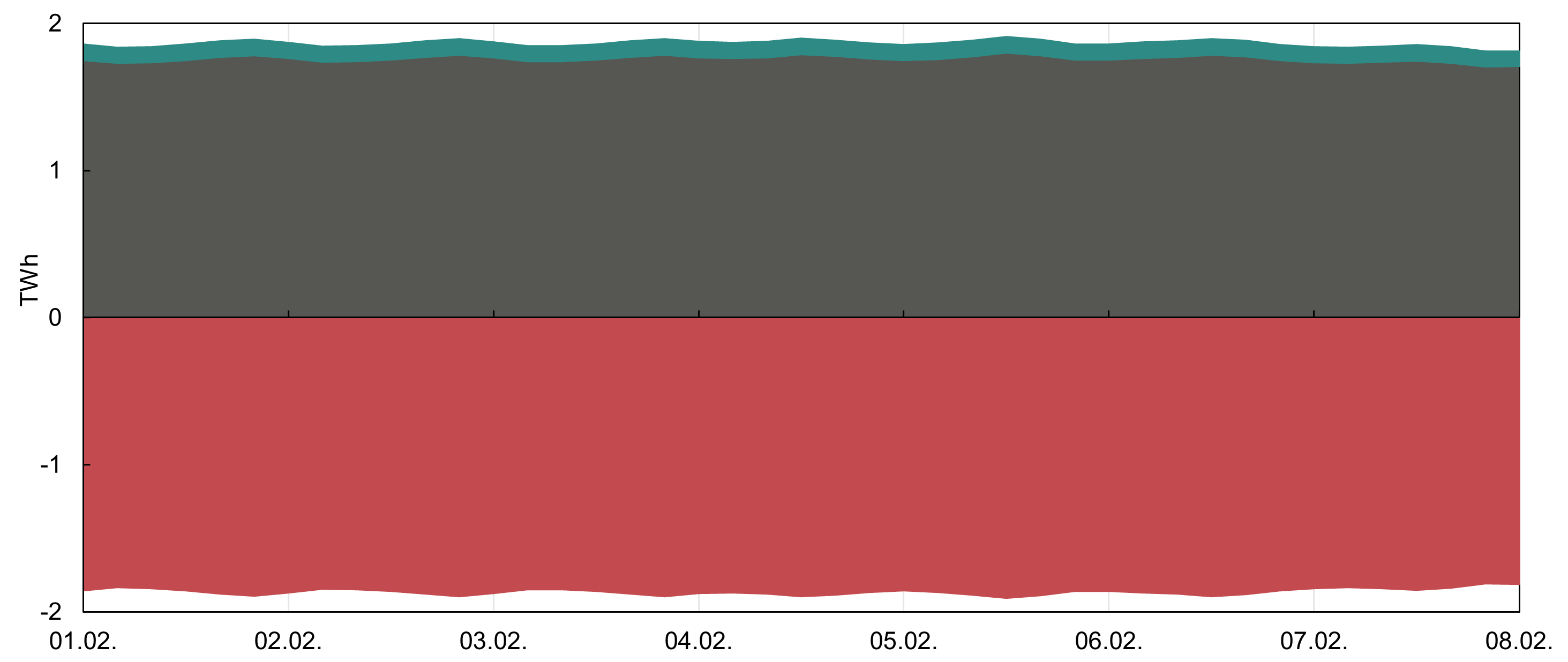} 
    \caption{Highest cost scenario (mean FOAK)}
    \label{fig:processHeat_low_FOAK}
    \end{subfigure}
    \begin{subfigure}{0.9\textwidth}
    \includegraphics[width=0.9\linewidth]{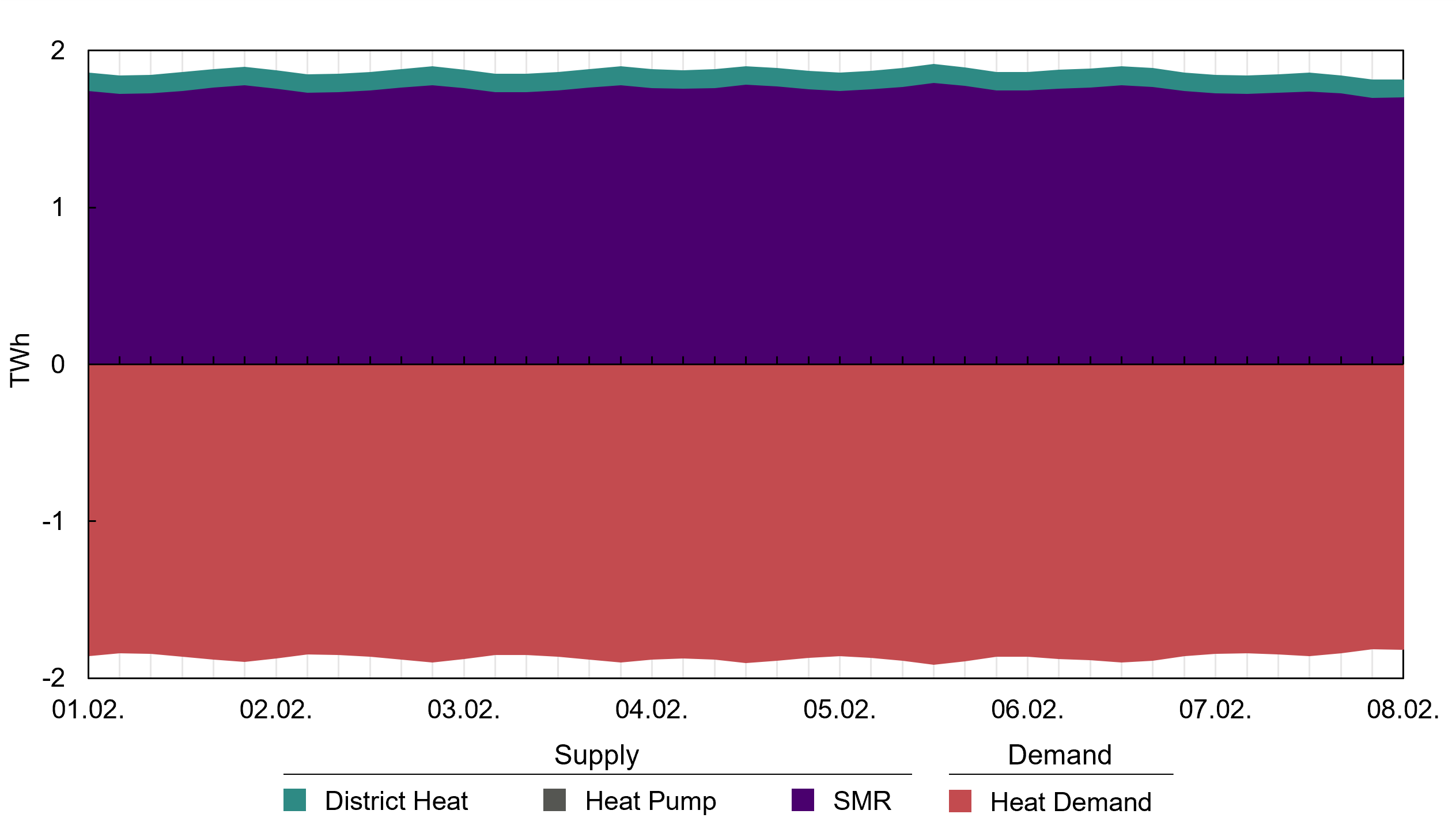}
    \caption{Least cost scenario (minimum NOAK)}
    \label{fig:processHeat_low_NOAK}
    \end{subfigure}
    \caption{Exemplary low-temperature process heat generation in France}
    \label{fig:timeSeries_processHeat_low_FR}
\end{figure}

\begin{figure}[h!]
\centering
    \begin{subfigure}{0.7\textwidth}
    \includegraphics[width=0.9\linewidth]{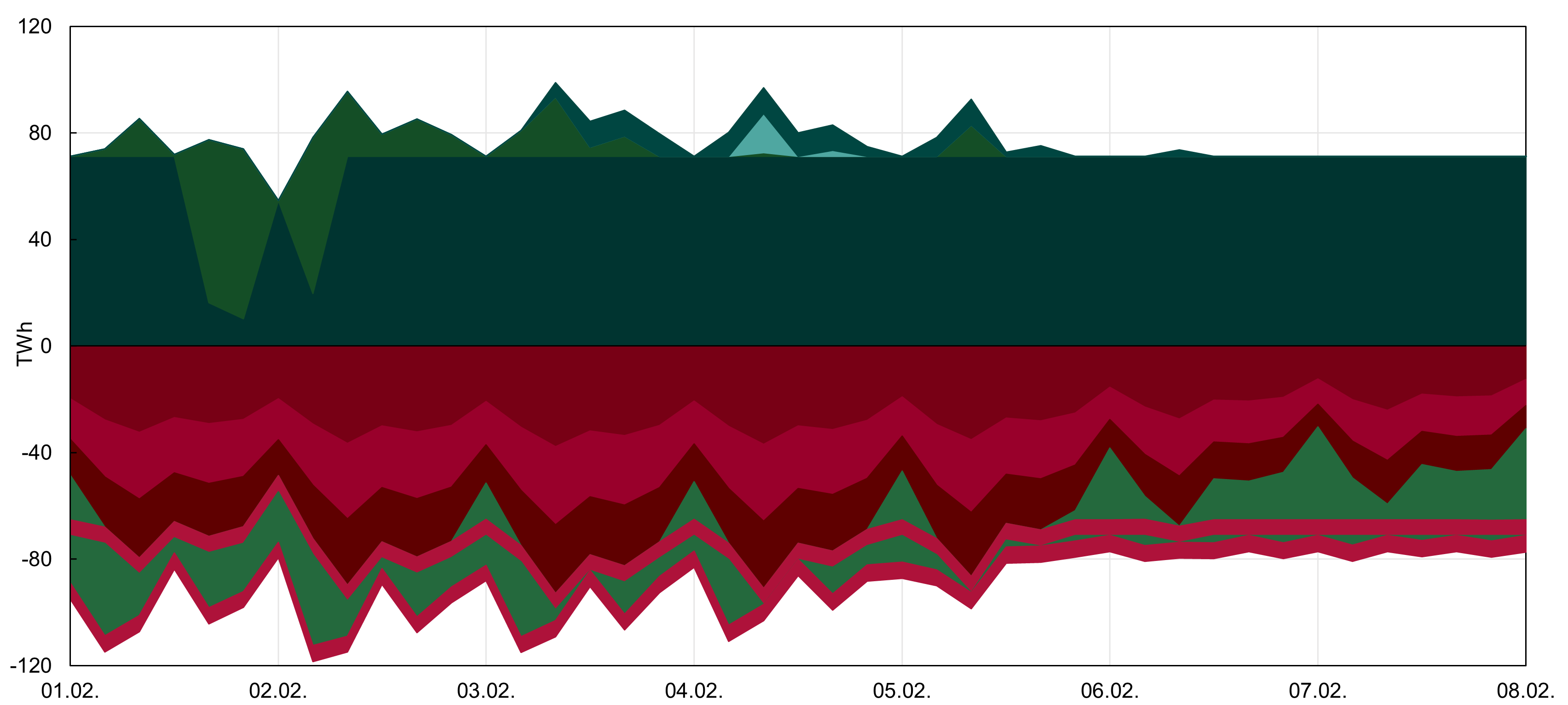} 
    \caption{Highest cost scenario (mean FOAK)}
    \label{fig:districtHeat_FOAK}
    \end{subfigure}
    \begin{subfigure}{0.7\textwidth}
    \includegraphics[width=0.9\linewidth]{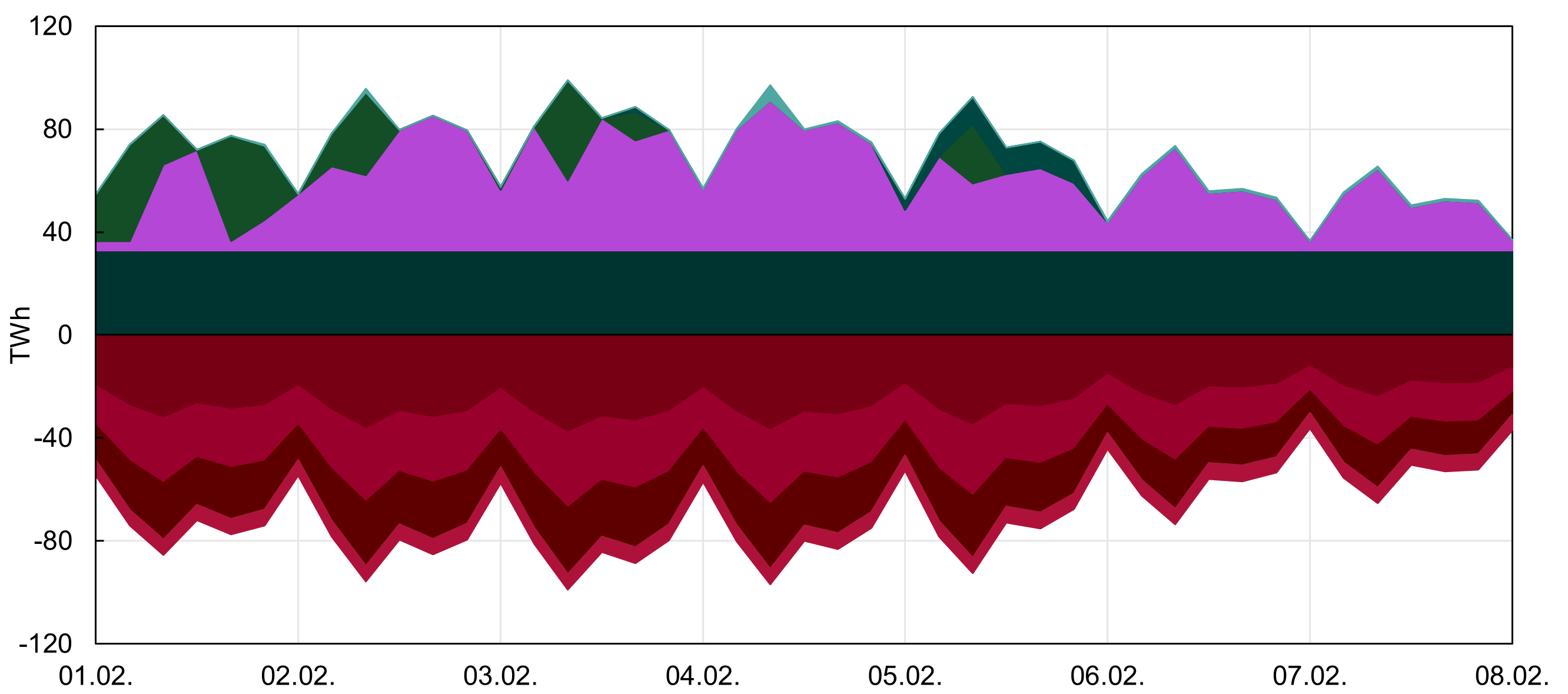}
    \caption{Second-to-last cost scenario}
    \label{fig:districtHeat_FOAK-8}
    \end{subfigure}
    \begin{subfigure}{0.7\textwidth}
    \includegraphics[width=0.9\linewidth]{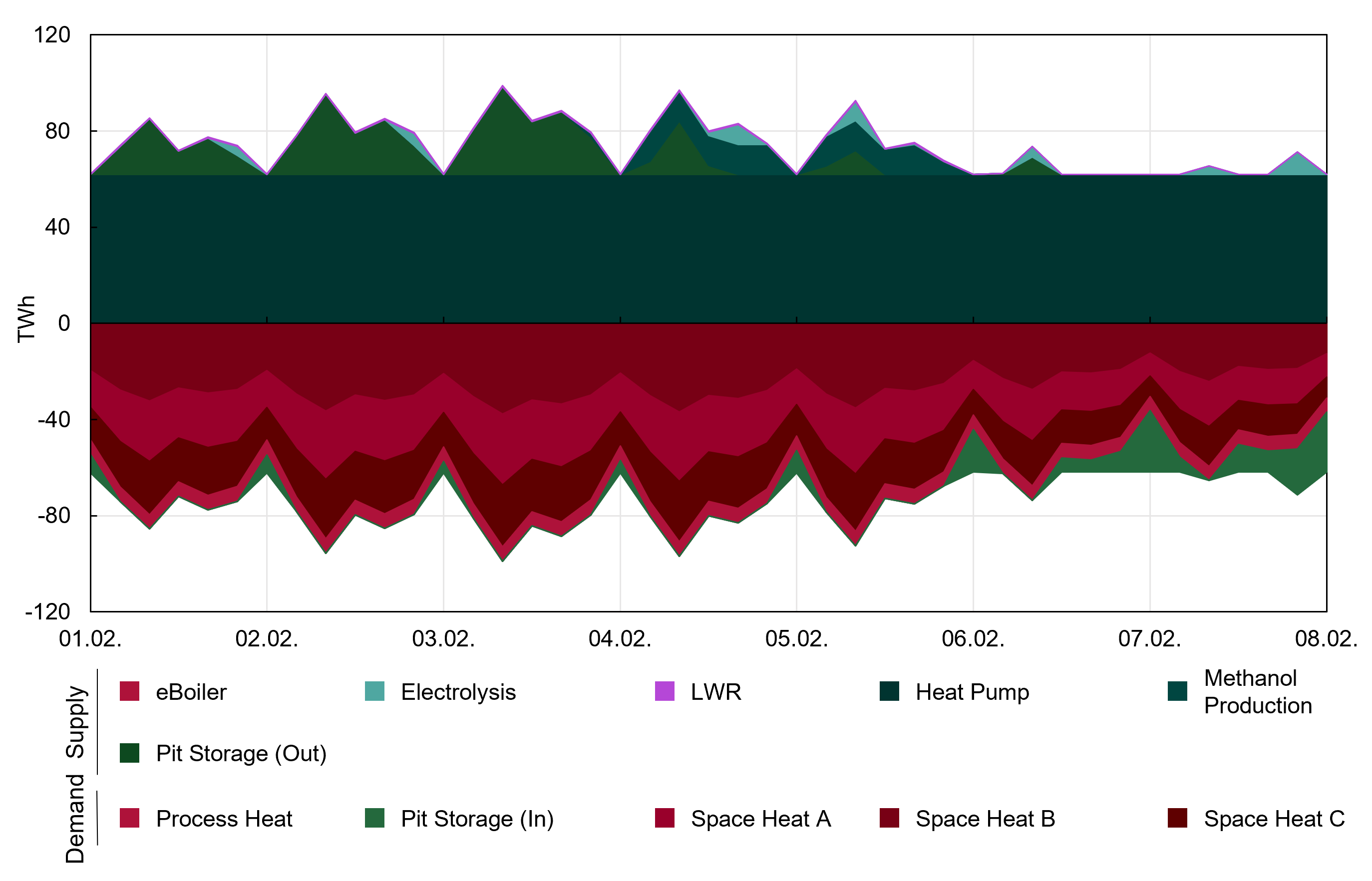} 
    \caption{Highest cost scenario (mean FOAK)}
    \label{fig:districtHeat_FOAK}
    \end{subfigure}
    \caption{Exemplary district heat generation in Germany}
    \label{fig:timeSeries_district_DE}
\end{figure}

\section{Energy Vectors and Technologies in the System Model}\label{secA4}

\begin{figure}[h]
 \centering
 \includegraphics[width=\linewidth]{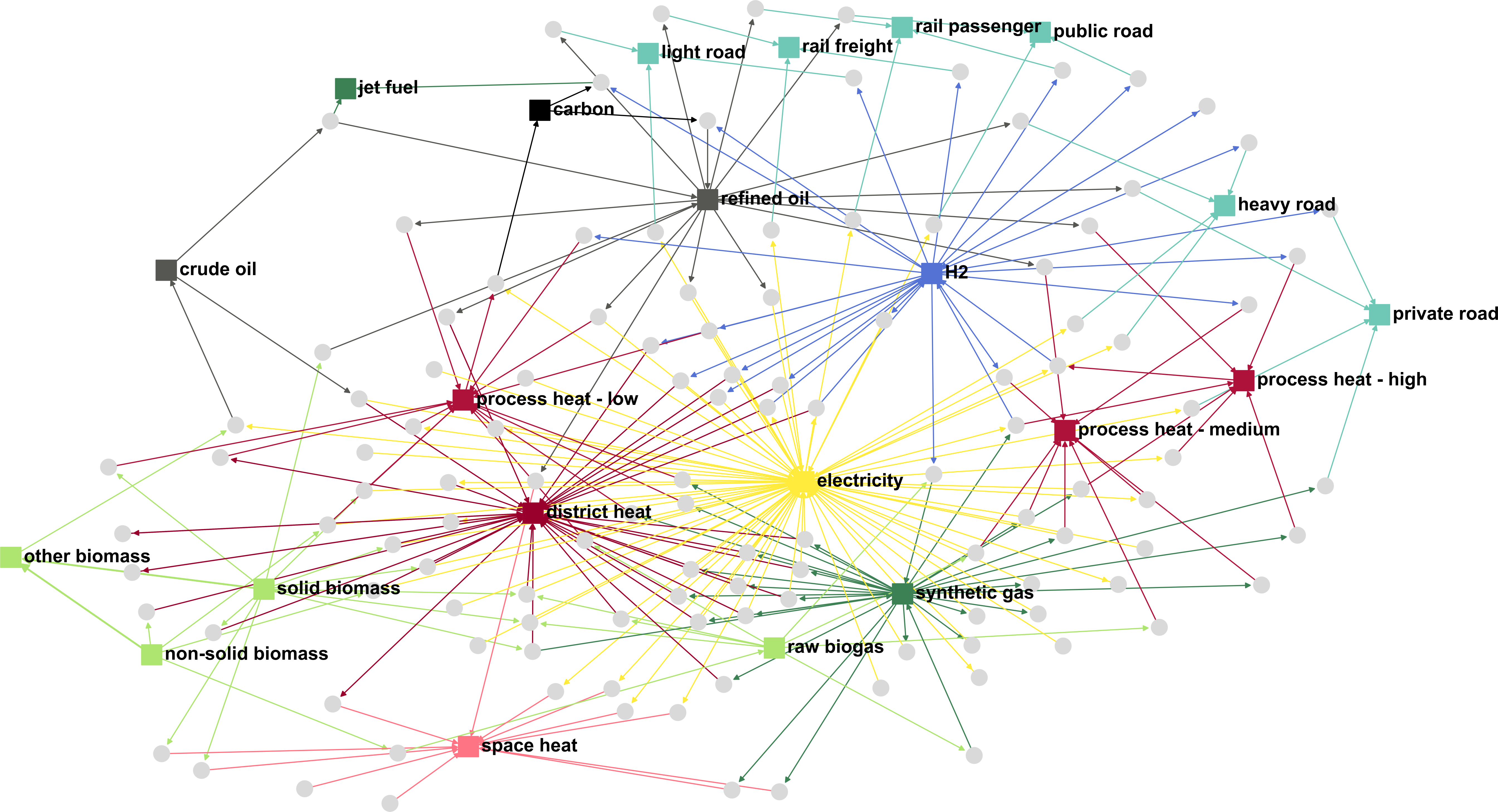}
 \caption{Full graph of model carriers and technologies based on \citet{goke_anymodjl_2021}.}
 \label{fig:all}
\end{figure}

The section gives a more detailed overview of the technology options and final demand categories considered in the model. Since the described features of the model are not novel or specific to this publication, the description is very similar to that of \citet{goke_flexible_2023}.
Overall, the model covers 22 energy carriers and 125 technologies, excluding the nuclear technologies we explicitly added for this paper. To illustrate the resulting complexity, Fig \ref{fig:all} Fig \ref{fig:all}. In this graph, the entering edges of technologies refer to input carriers, and the outgoing edges refer to outputs.

For primary supply, the model considers the following renewable technologies. Electricity generation from PV differentiates photovoltaic in open spaces, on roofs of residential buildings, and on roofs of industrial buildings; electricity generation from wind differentiates on- and offshore. All wind and PV technologies fluctuate, and their supply is contingent on an exogenous time-series of capacity factors. For hydro supply, the model considers run-of-river and reservoirs. Run-of-river fluctuates, like wind and PV. Reservoirs have exogenous and fluctuating inflows, but supply is dispatchable as long as the water in the reservoir is sufficient. For biomass, the most versatile energy source, the model distinguishes raw biogas, solid biomass, and non-solid biomass. Boilers or CHP plants can directly burn biomass to generate electricity, district heat, space heat, or process heat up to 500°C. Alternatively, various routes are available to convert biomass into synthetic fuels, like the gasification of solid biomass or upgrading raw biogas to synthetic methane and the liquefaction of solid and non-solid biomass to synthetic methanol. 
We use data from \citet{auer_development_2020} for the technical potential of wind and PV. To put these numbers into context, we compare them against the results of a meta-analysis on renewable potential in Europe in Tab. \ref{tab:ee_pot2} and Germany in Tab. \ref{tab:ee_pot} \citep{risch_potentials_2022, dupre_la_tour_photovoltaic_2023}. The material demonstrates that the assumed potential aligns with the meta-analyses' results. For Germany, capacities are close to the reference, assuming current legislation except for PV; here, we are more conservative. On a European level, our assumptions are close to or lower than the median of literature values.

\begin{table}
\caption{Comparison of renewable potential for Europe (excl. Norway and Balkan) in GW.} \label{tab:ee_pot2}
\centering
\begin{tabular}{lcccc}
\hline
&Own assumptions & \multicolumn{3}{c}{Literature values in} \\
 & based on & \multicolumn{3}{c}{ \citet{dupre_la_tour_photovoltaic_2023}}\\
 & \citet{auer_development_2020} & First quartile & Median & Third quartile\\
\hline
Wind, onshore & 4,508 & 1,518 & 3,421 & 7,489 \\
 Wind, offshore & 1,678 & 1,628 & 4,844 & 8,704\\
 PV, openspace & 2,793 & 1,252 & 4,235 & 8,625 \\
PV, rooftop & 1,558 & 610 & 952 & 1,665\\ \hline
\end{tabular}
\end{table}

\begin{table}
\caption{Comparison of renewable potential for Germany in GW.} \label{tab:ee_pot}
\centering
\begin{tabular}{lcccc}
\hline
 & Own assumption & Results for & \multicolumn{2}{c}{Literature values} \\
 & based on & current legislation &  \multicolumn{2}{c}{in \citet{risch_potentials_2022}}\\
 & \citet{auer_development_2020} & in \citet{risch_potentials_2022} & Lowest & Highest\\
 \hline
Wind, onshore & 386 & 385 & 68 & 1,188 \\
 Wind, offshore & 84 & 79 & 34 & 99.6\\
 PV, openspace & 301 & 456 & 90 & 1,285\\
 PV, rooftop & 177 & 492 & 43 & 746\\
 \hline
\end{tabular}
\end{table}
In the case of hydropower, we assume today's capacities are available without any investment due to their long technical lifetime but prohibit any expansion, assuming the available potential is fully utilized. The use of biomass in each country is subject to an upper energy limit of 1,081 TWh for the entire model, including waste for the production of biomass \citep{commission_of_the_european_union_joint_research_centre_institute_for_energy_and_transport_jrc-eu-times_2015}.

Besides the generation of primary supply, the model includes several technologies that provide secondary energy, which is the intermediate step from primary supply to final demand. Heat-pumps and electric boilers convert electricity into district heat; equally, each synthetic fuel, hydrogen, methane, or methanol, can fuel a different boiler technology to provide district heat. Alkali, solid-oxide, or proton exchange membrane electrolyzers utilize electricity to generate hydrogen while feeding waste heat into district heating networks. In addition to domestic production, the model can import hydrogen by ship at costs of 131.8 US-\$ per MWh and by pipeline from Morocco or Egypt at 90.7 and 86.8 US-\$ per MWh, respectively \citep{hampp_import_2023}. Vice versa, synthetic fuels can fuel different engines or turbines to generate electricity and district heat. In addition, there are several conversion pathways among the synthetic fuels: methane pyrolysis creates hydrogen from methane, methanation synthesizes methane from hydrogen (and raw biogas), and hydrogen-to-methanol conversion is possible but requires a carbon input supplied by direct air capture. 

Furthermore, the model can invest in storage systems for secondary energy carriers, as indicated by the arrows starting and ending at the same carrier. Electricity storage encompasses lithium-ion batteries, redox flow batteries, and pumped hydro storage. Short-term storage of district heat requires investment in water tanks and long-term storage in pit thermal storage. Caverns can store hydrogen and synthetic gas but depend on geological conditions, so the technical potential in each region is limited. For gas, the potential corresponds to today's gas storages; the potential for hydrogen caverns builds on \citet{caglayan_technical_2020}. In addition, investment into tanks for hydrogen storage is possible without any restrictions on potential. Still, investment costs per energy are substantially higher.

The right column of Fig. \ref{fig:overMod} lists the final energy demands the model considers. The final electricity demand corresponds to residential, service, and industry appliances. 
Next, there is a final demand for space and process heating. The model further splits process heat into three temperature levels: low temperature up to 100°C, medium temperature from 100°C to 500°C, and high temperature above 500°C. To cover the demand for space heat, the model can invest in solar thermal heating, boilers and heat-pumps fuelled by electricity, the connection to a district heating network, and boilers fuelled by biomass or synthetic fuels. The same options are available for low temperature heat, except for solar thermal heating; medium and high temperatures preclude heat-pumps and district heating. Even if a technology can provide a particular heat type, its capacity can still be constrained. The potential for district heating and ground-source heat-pumps in space heating reflects settlement structures. District heating can only supply heat in cities, towns, and suburbs; ground-source heat-pumps in rural areas and half the demand in towns and suburbs. In process heating, today's shares of district heating and electric boilers are an upper limit since it is indeterminate whether their temperature level is sufficient to supply a larger share of demand. Technologies for space heat and low-temperature heat utilize heat storage.

Finally, the model encompasses the demand for transport services. Passenger transport includes private road, public road, and public rail transport; freight includes heavy road, light road, and rail transport. Vehicle options for road transport include BEVs, fuel cells, and internal combustion engines using synthetic fuels. In addition, compressed natural gas vehicles are available for private passenger road transport and electric overhead lines for heavy freight transport. Rail transport can run on electricity, diesel, or fuel cell engines. Finally, there is a fixed demand for methanol for aviation and navigation.

\section{Spatial Resolution and Infrastructure in the System Model}\label{secA5}

This appendix provides the exact spatial resolutions and detailed modeling assumptions for exchange infrastructure. Since the described features of the model are not novel or specific to this publication, the description is very similar to the one in \citet{goke_flexible_2025}.

\begin{figure}
 \centering
  \includegraphics[scale=0.25]{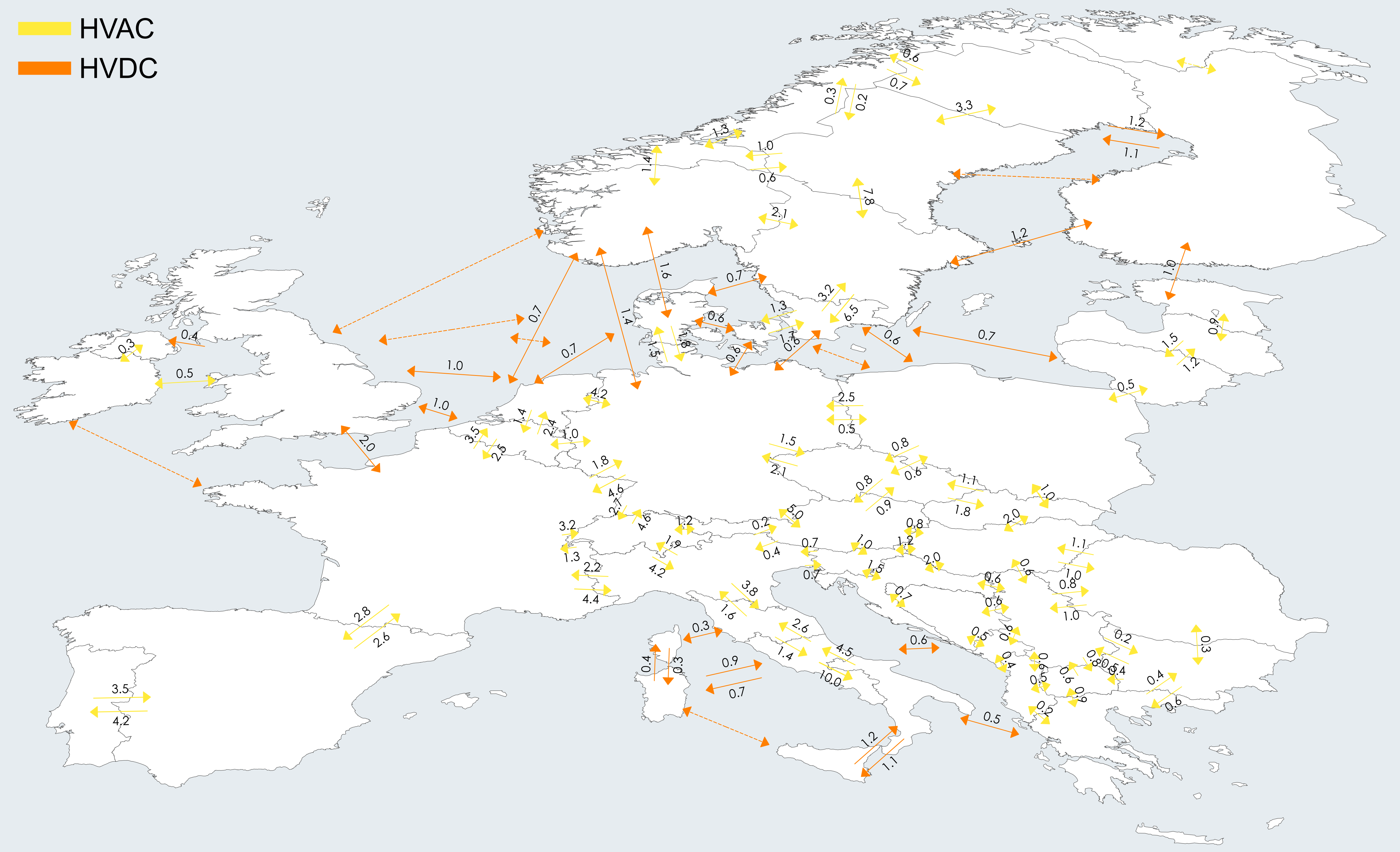}
 \caption{Spatial resolution and grid connections for electricity \citep{goke_flexible_2025}}
 \label{fig:startGrid1}
\end{figure}

The spatial resolution for electricity corresponds to the electricity market zones as shown in Fig. \ref{fig:startGrid1}. The exchange of electricity between zones requires HVAC (high-voltage alternating current) or HVDC (high-voltage direct current) lines, as indicated by the arrows in the figure. Whether HVAC or HVDC transmission connects two zones depends on the information provided by the European grid operator, but generally, HVDC is limited to long distances over sea \citet{entso-e_completing_2020}. The model can invest in new transmission lines; today's lines are available without investment, only incurring maintenance costs to reflect their long technical lifetime. The costs of new lines also include maintenance costs. The numbers in Fig. \ref{fig:startGrid1} specify today's capacities. The dotted arrows represent potential connections without any pre-existing capacity.

\begin{figure}[!htbp]
 \centering
\includegraphics[scale=0.13]{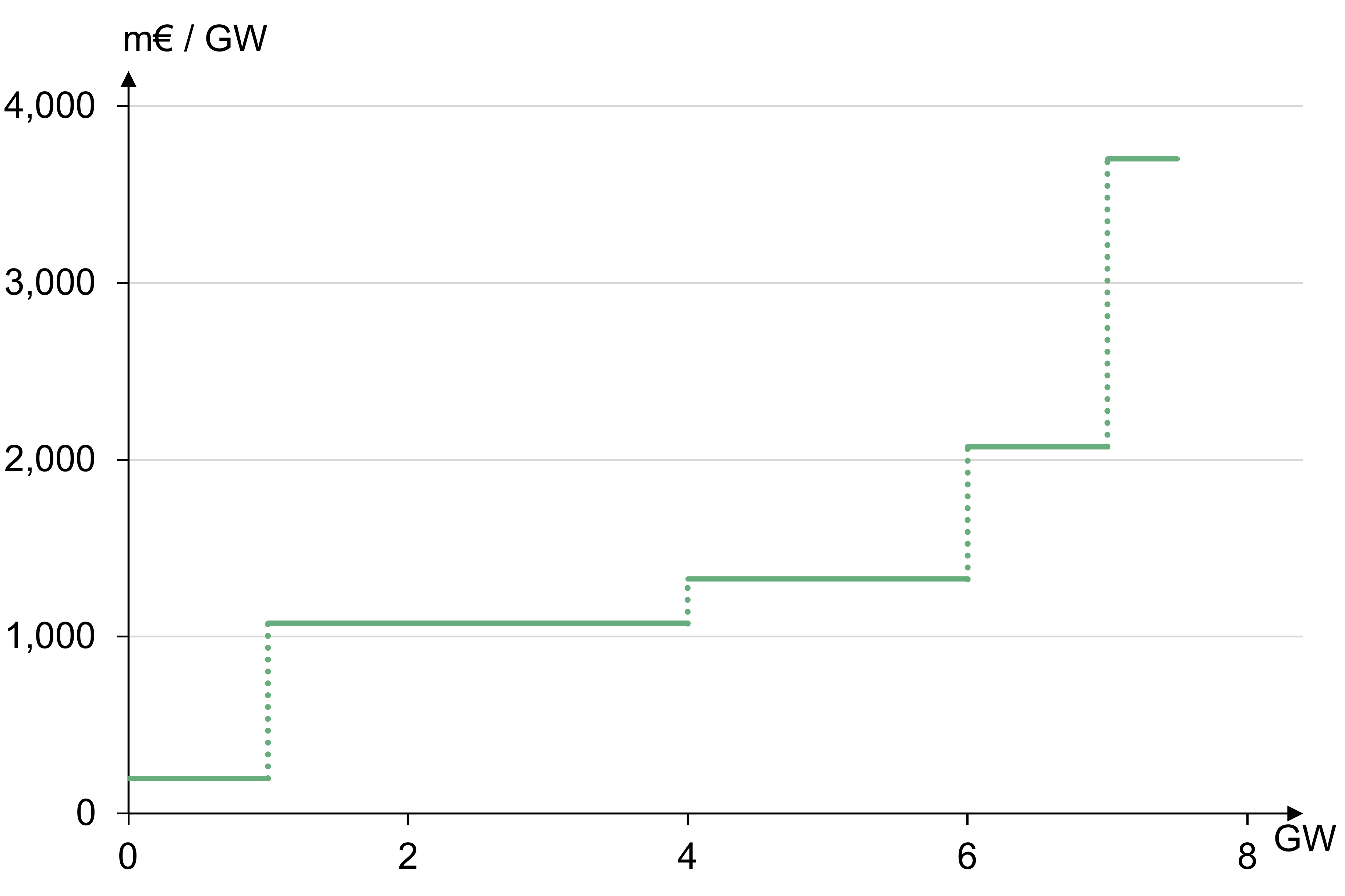}
 \caption{Potential-cost curve for expanding the electricity grid between Germany and the Netherlands \citep{goke_flexible_2025}}
 \label{fig:ntcExp}
\end{figure}

Electricity exchange within a market zone is unrestricted. This "copper plate" assumption neglects interzonal congestion but it allows building on specific ENTSO-E data listing potential expansion projects in the European electricity grid \citet{entso-e_completing_2020}. From this data, we obtain a potential-cost curve for each connection with increasing investment costs and an upper limit on expansion. For example, Fig. \ref{fig:ntcExp} shows this curve for the connection between Germany and the Netherlands. In this case, the specific investment costs rise from 200 to 3,700 million € per GW, and the total expansion limit is 7.5 GW. A more detailed spatial resolution cannot utilize this data and has to rely on potential and cost data that is not connection-specific and subject to substantial uncertainty; for instance, literature values for expansion costs vary by a factor of 5 for underground cables \citep{acer_unit_2023, 50hertz_kostenschatzungen_2021}.
We use a transport instead of flow formulation to model the grid operation since previous research found this simplification to be sufficiently accurate \citep[][cited by \citealp{goke_how_2023}]{neumann_assessments_2022}. In line with the same source, transmission losses amount to 5\% and 3\% per 1,000 km for HVAC and HVDC grids, respectively. The distance used is the distance between the geographic centers of each zone.

\begin{figure}
 \centering
  \includegraphics[scale=0.25]{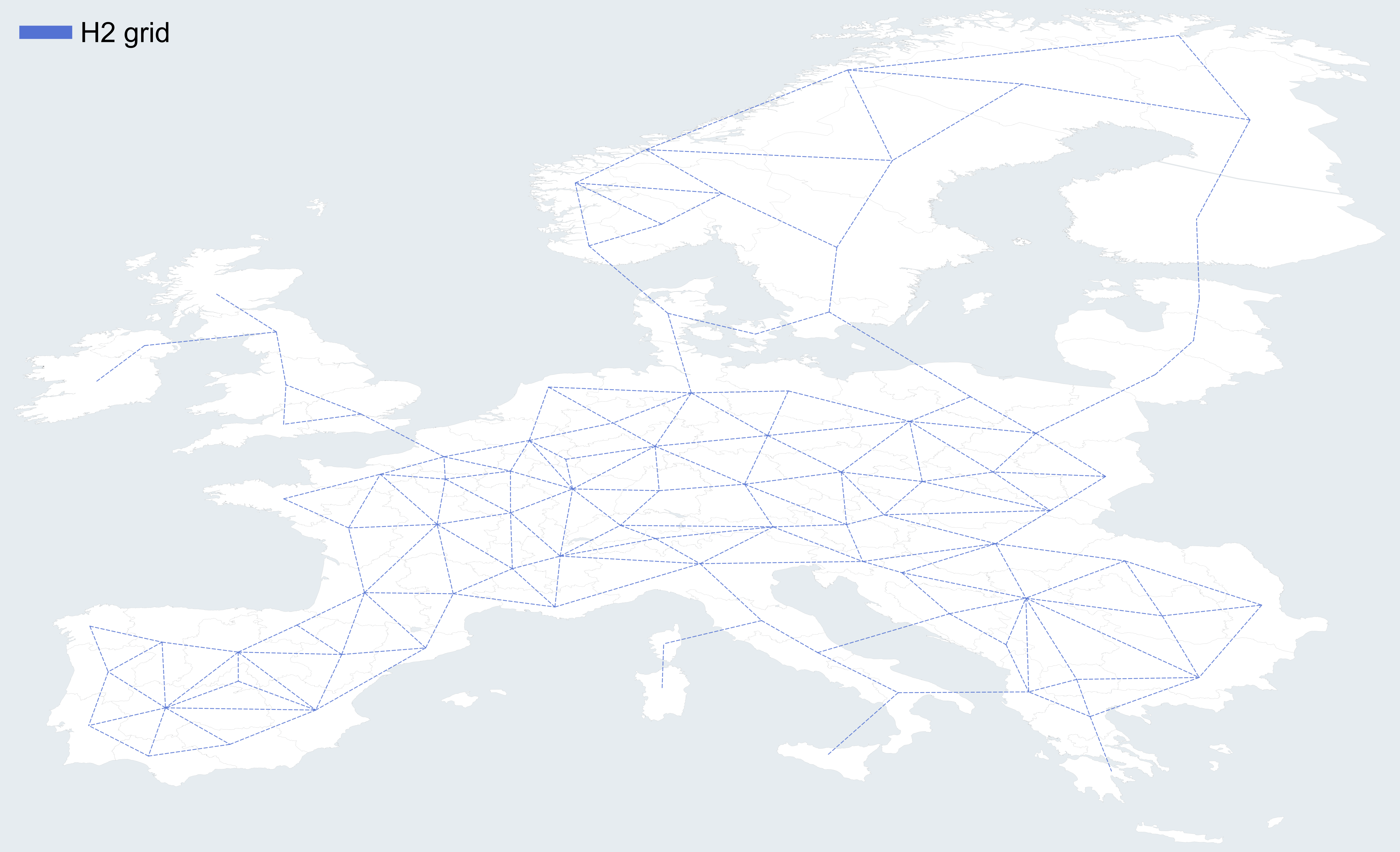}
 \caption{Spatial resolution and grid connections for hydrogen \citep{goke_flexible_2025}}
 \label{fig:startGrid2}
\end{figure}

For hydrogen exchange between clusters, the model can invest in pipelines indicated by the dotted lines in Fig. \ref{fig:startGrid2}. These pipelines are subject to costs of 0.4 million EUR per GW and km and energy losses of 2.44\% per 1,000 km \citep{danish_energy_agency_technology_2022}. The distance between the geographic center of clusters serves as an estimate for pipeline length. The exchange of methanol and biomass between countries is possible but does not require grid infrastructure. Instead, it uses trucks and only incurs variable costs based on \citet{commission_of_the_european_union_joint_research_centre_institute_for_energy_and_transport_jrc-eu-times_2015}. 
